\documentclass[conference]{IEEEtran}

\usepackage[square,numbers,sort&compress]{natbib}
\usepackage{hyperref} % 
\usepackage% 
{hyperref} % 
\hypersetup{
    colorlinks,
    linkcolor={blue!80!black},% 
    citecolor={green!50!black},
    urlcolor={blue!80!black}
}

\usepackage{amsthm}
\usepackage{amsmath}
\usepackage{mathrsfs}
\usepackage{amssymb} 
\usepackage{stackrel}
\usepackage{mathtools}
\usepackage{nicefrac}

\usepackage{physics}

\usepackage{verbatim}
\usepackage{enumerate}

\usepackage[T1]{fontenc}
\usepackage{bbm} 
\usepackage{bbding}
\usepackage{keystroke}
\usepackage{pifont}
\usepackage{relsize} % 
\usepackage{bm}

\usepackage{graphicx}
\usepackage[table]{xcolor}
\usepackage{tikz}
\usetikzlibrary{shapes, arrows.meta, positioning}
\usepackage{pgfplots}
\pgfplotsset{compat=1.18}
\usetikzlibrary{calc}
\usepackage{tabularx}
\usetikzlibrary{arrows,shapes}

\definecolor{sourcecolor}{HTML}{9B2C2C}
\definecolor{relaycolor}{HTML}{6F42C1}
\definecolor{destcolor}{HTML}{1A5FB4}
\definecolor{wardencolor}{HTML}{00703C}
\definecolor{statecolor}{HTML}{19764A}

\usepackage{balance}

\renewcommand{\trace}{\mathrm{Tr}}

\newcommand{\identity}{\mathbbm{1}}

\newcommand{\markovC}[1]{% 
\begin{tikzpicture}[#1]% 
\draw (0,0.3ex) -- (1ex,0.3ex);% 
\draw (0.5ex,0.3ex) circle (0.2ex);
\draw[white] (0.2ex,0) -- (0.5ex,0);% 
\end{tikzpicture}% 
}
\newcommand{\Cbar}{\markovC{scale=2}}

\theoremstyle{remark}	\newtheorem{theorem}{Theorem}
\theoremstyle{remark}	\newtheorem{lemma}[theorem]{Lemma}
\theoremstyle{remark}	
\theoremstyle{remark}	\newtheorem{proposition}[theorem]{Proposition}
\theoremstyle{remark} \newtheorem{definition}{Definition}
\theoremstyle{remark} \newtheorem{remark}{Remark}
\theoremstyle{remark}

\title{Can Quantum Entanglement Increase the Capacity of a Classical Relay Channel?}
\author{% 
  \IEEEauthorblockN{Elsa Mizrahi}
  \IEEEauthorblockA{Department of Electrical Engineering \\
                    Télécom Paris, Institut Polytechnique de Paris\\
                    Palaiseau, France\\
                    Email: {elsa.mizrahi}@telecom-paris.fr}
  \and
  \IEEEauthorblockN{Uzi Pereg}
  \IEEEauthorblockA{ECE \& Helen Diller Quantum Center\\ 
                    Technion -- Israel Institute of Technology\\
                    Haifa 3200003, Israel\\
                    Email:  uzipereg@technion.ac.il}
}

\date{\today}

\begin{document}

\maketitle

\begin{abstract}
We show positive and negative results on classical relay channels with entanglement assistance, revealing a sharp contrast between strictly causal and causal relay encoding.
 First, we consider van der Meulen's setting of a unit-delay relay, with strictly causal encoding. We show that the standard cutset upper bound continues to hold with entanglement assistance, and deduce that entanglement does not increase the capacity for several classes of relay channels. Second, we consider a zero-delay relay, where the encoding is causal. We present an example in which entanglement enables rates that exceed the cutset upper bound by at least $ 12.5\%$. Namely, entanglement shared between the source and relay significantly increases the capacity of a classical zero-delay relay channel.
\end{abstract}

\maketitle

\section{Introduction}

Relays can extend communication range by helping a sender transmit information to a distant destination. Figure~\ref{fig:Relay_schematic} illustrates the basic three-terminal relay channel, consisting of a sender, relay, and destination: 
\begin{figure}[bth]

    \centering
    
\begin{tikzpicture}[scale=0.5,
    node distance=1.5cm and 1.8cm,
    >=latex,
    sum/.style={draw, circle, inner sep=0pt, minimum size=2mm},
    box/.style={draw, align=center}
]

\node [sum,thick,draw=sourcecolor,fill=sourcecolor!90] at (3,3) (v1) {};
\node [sum,thick,draw=destcolor,fill=destcolor!90] at (9,3) (v2) {};
\node [sum,thick,draw=relaycolor,fill=relaycolor!90] at (6,5) (v3) {};

    \begin{scope}[line width=1pt]

             \draw[->,thick] (v1) node[left,xshift=-0.25cm,yshift=0.025cm]{ {sender}} -- (v2) node[right,xshift=0.25cm,yshift=0.025cm]{ {destination}};
             \draw[->,thick] (v3)  -- (v2) node[right,xshift=0.25cm,yshift=0.025cm]{};
		\draw[->,thick] (v1) -- (v3) node[right,xshift=0.25cm,yshift=0.075cm]{ {relay}};

 \end{scope}

\end{tikzpicture}

\caption{\label{fig:Relay_schematic}Relay schematic.}
\end{figure}
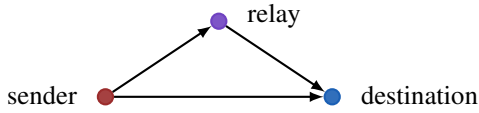
\\% 
Figure~\ref{Fig:relay_diag} shows the associated signals% 
, 
where
$X$ denotes the sender's transmission, $Y_r$ the relay's observation, $X_r$ the relay's transmission, and $Y$ the destination's observation% 
:
\begin{figure}[bht]

\center
\begin{tikzpicture}[scale=0.5,
    node distance=1.5cm and 1.8cm,
    >=latex,
    sum/.style={draw, circle, inner sep=0pt, minimum size=2mm},
    box/.style={draw, align=center}
]

    \tikzstyle{block} = [draw, shape=rectangle, minimum height=6em, minimum width=3cm, node distance=4cm, line width=1pt]
    \tikzstyle{sum} = [draw, shape=circle,node distance=4cm, line width=1pt, minimum width=0.25em]
     \tikzstyle{mult} = [draw, shape=circle, node distance=3cm, line width=1pt, minimum width=2.em]
 \tikzstyle{branch}=[fill,shape=circle,minimum size=4cm,inner sep=0pt]

\node [draw,thick,shape=circle,inner sep=1pt, minimum size=0.575cm,sourcecolor] at (3,3) (A) {\footnotesize{$\!X$}};
\node [draw,thick,shape=circle,inner sep=1pt, minimum size=0.575cm,destcolor] at (9,3) (B) {\footnotesize{$Y$}};

\node [draw,thick,shape=circle,inner sep=1pt, minimum size=0.575cm,relaycolor] at (5.4,5.5) (E) {\footnotesize{$Y_r$}};

\node [draw,thick,shape=circle,inner sep=1pt, minimum size=0.575cm,relaycolor] at (6.6,5.5) (D) {\footnotesize{$X_r$}};
    
\node  at (1,3) (v1i) {
{{sender}}};
\node  at (11.75,3) (v2i)  {% 
{{destination}}};
\node  at(8,6) (v3i) {
{{relay}}};

    \begin{scope}[line width=1pt]

             \draw[->,thick] (A) -- (B);
		\draw[->,thick] (A) -- (E);
		\draw[->,thick] (D) --(B);

 \end{scope}

\end{tikzpicture}

\caption{\label{Fig:relay_diag}Relay channel.}
\end{figure}
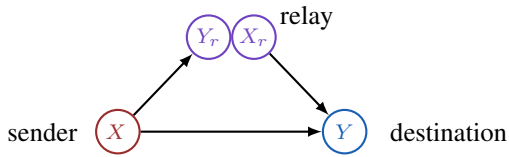

\begin{figure}[bh]
\center
\begin{tikzpicture}[every node/.style={scale=1}]
\tikzstyle{myedgestyle} = [-triangle 60]

    \tikzstyle{block} = [draw, shape=rectangle, minimum height=6em, minimum width=1cm, node distance=2cm, line width=1pt]
    \tikzstyle{sum} = [draw, shape=circle, node distance=2cm, line width=1pt, minimum width=1.em]
     \tikzstyle{mult} = [draw, shape=circle, node distance=1cm, line width=1pt, minimum width=1.em]
 \tikzstyle{branch}=[fill,shape=circle,minimum size=4cm,inner sep=0pt]
\tikzset{meter/.append style={draw, inner sep=10,  font=\vphantom{A}, minimum width=30, line width=.8,
 path picture={\draw[black] ([shift={(.1,.3)}]path picture bounding box.south west) to[bend left=50] ([shift={(-.1,.3)}]path picture bounding box.south east);\draw[black,-latex] ([shift={(0,.1)}]path picture bounding box.south) -- ([shift={(.3,-.1)}]path picture bounding box.north);}}}

    \begin{scope}[line width=0.5pt]

			\draw[->] (-4,3) -- (-3,3) node[midway,above] {$m$};

           \draw[->] (-2,3) -- (-1,3)     node[midway,above] {$X_i$};

		\draw  (-3,3.5) rectangle (-2,2.5) node[midway] 
        {Enc};
		
		\draw  (-1,3.5) rectangle (1.5,2.5) node[midway] { $P_{YY_r|XX_r}$};

		\draw  (-1,5.5) rectangle (1.5,4.5) node[midway] { Relay Encoder} ;

        \draw[->] (-0.5,3.5) -- (-0.5,4.5) node[midway,left,] {\textcolor{relaycolor!20!red}{$Y_r^{i-1}$}$/$\textcolor{relaycolor!20!blue}{$Y_r^{i}$}};
        \draw[->] (1,4.5) -- (1,3.5) node[midway,right] {$X_{r,i}$};

	    \draw[->] (1.5,3) -- (2.5,3) node[midway,above] {$Y_i$};

		\draw  (2.5,3.5) rectangle (3.5,2.5) node[midway] 
        {Dec};
        
        \draw[->] (3.5,3) -- (4.5,3) node[midway,above] {$\hat{m} $}; 

\end{scope}

\end{tikzpicture}
\caption{% 
Classical relay encoding. Alice's encoder maps message $m$ to a codeword $X^n$,  and transmits $X_i$ at time $i$. A unit-delay relay chooses its transmission $X_{r,i}$ using only past observations $Y_r^{i-1}$, i.e., the relay encoding is strictly causal (red). A zero-delay relay may also use the current observation $Y_{r,i}$, i.e., the encoding is causal (blue). Bob observes $Y_i$. After receiving $Y^n$, he produces an estimate $\hat m$ of Alice's original message.
\label{Figure:Classical_Relay}}
\end{figure}
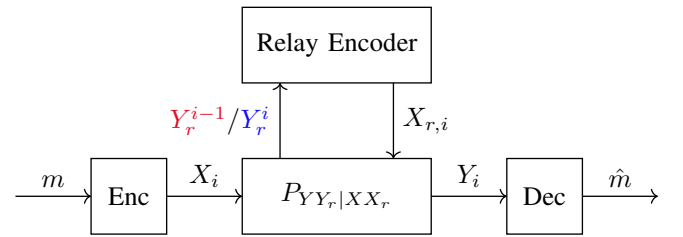

System synchronization determines the order in which the relay receives and transmits within each time frame \cite{Kramer:08n}. 
If the relay transmits before receiving, its new observation can only affect subsequent transmissions. If it receives before transmitting, it can use that observation in the same channel use. These two operating modes correspond to different \emph{encoding delays}. In the unit-delay model introduced by van der Meulen \cite{vanderMeulen:71p}, $X_{r,i}$ can depend only on past observations $Y_r^{i-1}$. In the zero-delay model studied by El Gamal and Hassanpour \cite{ElGamalHassanpour2005}, $X_{r,i}$ may also depend on the current observation $Y_{r,i}$.
These models of operation are also known as strictly-causal and causal relay encodings, respectively \cite{ElGamalKim:11b},
as illustrated in Figure~\ref{Figure:Classical_Relay}.
This difference in synchronization can strictly increase capacity \cite{ElGamalHassanpourMammen:07p}.

The role of shared entanglement in classical networks has become an active area of research.
 Leditzky et al. \cite{LeditzkyAlhejjiLevinSmith:20p} demonstrated that entanglement between senders can increase their
 sum-rate capacity, for a two-user  classical multiple access channel (MAC). 
 The result % 
 was later improved in  \cite{SeshadriLeditzkySiddhuSmith}, showing an increase of at least $4.7\%$ in the sum-rate capacity.
 Pereg et al. \cite{PeregDeppeBoche:21p2} showed that entanglement does \emph{not} increase achievable rates in the mirrored setting of a broadcast channel with entangled receivers. Other multi-user settings include 
 interference channels~\cite{HawellekMohanAghaeeDeppe:25c}, 
 source coding~\cite{SalehiShiraniPradhan:25c},
 network coding~\cite{YNPRLB:26p},
 privacy~\cite{ZivarifardWang:26p}, and non-signaling correlations \cite{FawziFermé:22,FawziFerme:24p,% 
YaoJafar:26a2}.
The effect of channel uncertainty  is further studied in \cite{Noetzel:20p,YaoJafar:26a}. % 
Here, however, we focus on single-user communication.

Quantum correlations can be examined through the lens of nonlocal games \cite{scarani2019bell}. In such a game, a referee sends a question to each player, and the players return answers without communicating. They win if the questions and answers satisfy a prescribed condition. A game is called pseudo-telepathic if players sharing entanglement can win with probability one, while no classical strategy can do so. The magic-square game is an example \cite{brassard2005quantum}. Leditzky et al. \cite{LeditzkyAlhejjiLevinSmith:20p} defined a class of  MACs based on nonlocal games: a channel in this class is essentially noiseless when its inputs constitute a winning play and injects noise otherwise. Entanglement thus helps the senders coordinate their inputs to satisfy the winning condition.

In this work, we establish positive and negative results on classical relay channels with entanglement assistance, revealing a sharp contrast between strictly causal and causal relay encoding.
 First, we consider van der Meulen's setting of a unit-delay relay, with strictly causal encoding. We show that the standard cutset upper bound continues to hold with entanglement assistance, and deduce that entanglement does not increase the capacity for several classes of relay channels.
 Second, we consider a zero-delay relay, where the encoding is causal. We construct a channel based on the magic-square game
and show that entanglement enables rates that exceed the cutset upper bound by at least $ 12.5\%$. Namely, entanglement shared between the sender and relay significantly increases the capacity of a classical zero-delay relay channel.

To establish the separation in the zero-delay example, we bound the rates achievable without entanglement. Our converse is inspired by Seshadri et al. \cite{SeshadriLeditzkySiddhuSmith},
who study game-based MACs and translate locality into optimization constraints.
Our analysis further exploits the specific structure of the magic-square game and captures the parity relations among classical responses. % 

The paper is organized as follows. In Section~\ref{Section:Notation}, we provide our notation conventions. Section~\ref{Section:Definitions} describes the relay communication model, for both the unit-delay and zero-delay relay channels. In section~\ref{Sec:Main_SC}, we present our negative results on the unit-delay relay channel, and in Section~\ref{Sec:Main_Causal} we demonstrate the entanglement advantage for the zero-delay relay channel. Section~\ref{Section:Summary} concludes with a summary and discussion of future research directions. The Appendices are dedicated for proofs of the technical lemmas in our analysis. % 

\section{Basic Notation}
\label{Section:Notation}
 We use standard notation conventions \cite{Wilde:17b}. All logarithms are to base two. 
 A discrete random variable $X$ takes value in a finite alphabet $\mathcal{X}$ and has probability mass function (PMF) $p_X$.
  The associated entropy is $H(p_X)\equiv -\sum_{x\in\mathcal{X}} p_X(x)\log p_X(x)$, with the convention $0\log 0=0$.
Denote   the probability simplex on $\mathcal{X}$ by $\mathbb{P}_{\abs{\mathcal{X}}}$.
 Let
 $x^i=(x_1,x_2,\ldots,x_i)$ denote a sequence of symbols from $\mathcal{X}$. The corresponding random sequence and its PMF are denoted by $X^i$ and $p_{X^i}$. 
  For every positive integer $d$, let
 $[d]\equiv \{1,\ldots,d\}$. A vector of length $d$ is denoted by $\mathbf{v}=\begin{pmatrix}
     v_1&v_2&\ldots&v_d
 \end{pmatrix}^T$, and its $\ell^2$-norm  by $\norm{\mathbf{v}}_2$.

 Given a joint PMF $p_{XY}$ on $\mathcal{X}\times\mathcal{Y}$, the classical mutual information is defined as 
$I(X;Y)=H(p_X)+H(p_Y)-H(p_{XY})$.
The conditional entropy is defined by
$H(Y|X)=\sum_{x\,:\; p_X(x)>0}\,p_X(x) H(Y|X=x)$, where 
$H(Y|X=x)=-\sum_{y\in\mathcal{Y}} p_{Y|X}(y|x)\log p_{Y|X}(y|x)$.
According to the entropy chain rule, $H(Y|X)=H(XY)-H(X)$. 
Similarly, the conditional mutual information
$I(X;Y|Z)=H(X|Z)+H(Y|Z)-H(XY|C)$ and satisfies
$I(X;Y|Z)=\sum_{z\,:\; p_Z(z)>0}\,p_Z(z) I(X;Y|Z=z)$.

A quantum state is described by a density operator $\rho$ on a finite-dimensional  
Hilbert space $\mathcal{H}$. 
 Denote the set of density operators on $\mathcal{H}$ by $\mathfrak{D}(\mathcal{H})$.
The von Neumann entropy is $H(\rho)\equiv -\trace\left[ \rho\log\rho \right]$. 
A generalized measurement is described by operators $\{\mathsf{K}_x\}_{x\in\mathcal{X}}$ satisfying $\sum_{x\in\mathcal{X}}\mathsf{K}_x^\dagger \mathsf{K}_x=\identity$, where $\identity$ is the identity operator.  
 The probability of outcome $x$ % 
 is $p_X(x)=\trace(\mathsf{K}_x^\dagger \mathsf{K}_x \rho)$.

 For a bipartite state $\rho_{AB}\in \mathfrak{D}( 
\mathcal{H}_A\otimes \mathcal{H}_B)$, let $\rho_A$ and $\rho_B$ denote its reduced states. The quantum mutual information and conditional entropy are  
$I(A;B)_\rho=H(\rho_A)+H(\rho_B)-H(\rho_{AB})$ and
$H(A|B)_\rho=H(\rho_{AB})-H(\rho_B)$. For a tripartite state $\rho_{ABC}$,  the conditional mutual information is
$I(A;B|C)_\rho% 
$.

\section{Coding With Entangled Transmitters}% 
\label{Section:Definitions}

\begin{figure*}[tb]
\center

\begin{tikzpicture}[every node/.style={scale=1}]
\tikzstyle{myedgestyle} = [-triangle 60]

    \tikzstyle{block} = [draw, shape=rectangle, minimum height=6em, minimum width=1cm, node distance=2cm, line width=1pt]
    \tikzstyle{sum} = [draw, shape=circle, node distance=2cm, line width=1pt, minimum width=1.em]
     \tikzstyle{mult} = [draw, shape=circle, node distance=1cm, line width=1pt, minimum width=1.em]
 \tikzstyle{branch}=[fill,shape=circle,minimum size=4cm,inner sep=0pt]
\tikzset{meter/.append style={draw, inner sep=10,  font=\vphantom{A}, minimum width=30, line width=.8,
 path picture={\draw[black] ([shift={(.1,.3)}]path picture bounding box.south west) to[bend left=50] ([shift={(-.1,.3)}]path picture bounding box.south east);\draw[black,-latex] ([shift={(0,.1)}]path picture bounding box.south) -- ([shift={(.3,-.1)}]path picture bounding box.north);}}}

    \begin{scope}[line width=0.5pt]

            \draw[red] (-4,3) -- (-3,3) node[midway,above] {$A$};
            \draw[red] (-4,3) -- (-5,4) node[left] {\textcolor{black}{$\Psi$}};

            \draw[blue] (-4,5) -- (-3,5) node[midway,above] {$G_0$};
            \draw[blue] (-3,5) -- (-1,5) node[midway,above] {};
            \draw[blue] (-4,5) -- (-5,4) node[midway,above] {};

			\draw[->] (-2.5,2) node[below] {$m$} -- (-2.5,2.5) ;

           \draw[->] (-2,3) -- (-1,3)     node[midway,above] {$X_i$};

		\draw  (-3,3.5) rectangle (-2,2.5) node[midway] 
        {Enc};
		
		\draw  (-1,3.5) rectangle (1.5,2.5) node[midway] { $P_{YY_r|XX_r}$};

		\draw  (-1,5.5) rectangle (1.5,4.5) node[midway] { Relay Encoder} ;

        \draw[->] (-0.5,3.5) -- (-0.5,4.5) node[midway,left,yshift=0.2cm] {$Y_{r,i-1}$};
        \draw[->] (1,4.5) -- (1,3.5) node[midway,right] {$X_{r,i}$};

	    \draw[->] (1.5,3) -- (2.5,3) node[midway,above] {$Y_i$};

		\draw  (2.5,3.5) rectangle (3.5,2.5) node[midway] 
        {Dec};
        
        \draw[->] (3.5,3) -- (4.5,3) node[midway,above] {$\hat{m} $};

        \node[left,xshift=-0.15cm] at (-0.5,3.75) {\footnotesize{Delay}};

\end{scope}

 \begin{scope}[shift={(-0.5,3.725)},scale=0.85]
 \clip (-0.2,0) rectangle (0.2,0.2);
    \draw[fill=white] (0,0) circle(0.2);
    \draw (-0.2,0) -- (0.2,0);
\end{scope}

\end{tikzpicture}
\caption{Unit-delay relay with entangled transmitters.}
\label{Figure:Strictly_causal_encoding}
\end{figure*}
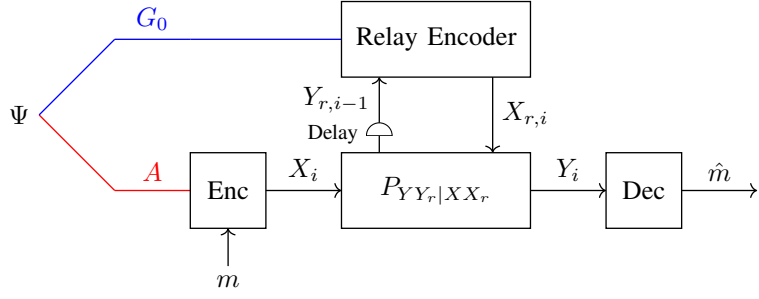

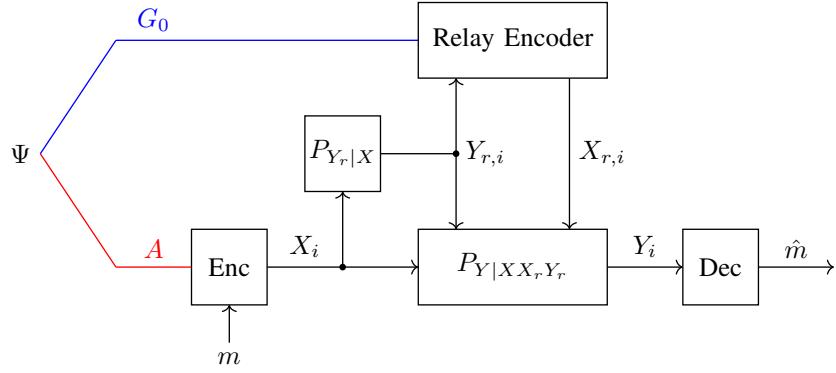
\begin{figure*}[tb]
\center

\begin{tikzpicture}[every node/.style={scale=1}]
\tikzstyle{myedgestyle} = [-triangle 60]

    \tikzstyle{block} = [draw, shape=rectangle, minimum height=6em, minimum width=1cm, node distance=2cm, line width=1pt]
    \tikzstyle{sum} = [draw, shape=circle, node distance=2cm, line width=1pt, minimum width=1.em]
     \tikzstyle{mult} = [draw, shape=circle, node distance=1cm, line width=1pt, minimum width=1.em]
 \tikzstyle{branch}=[fill,shape=circle,minimum size=4cm,inner sep=0pt]
\tikzset{meter/.append style={draw, inner sep=10,  font=\vphantom{A}, minimum width=30, line width=.8,
 path picture={\draw[black] ([shift={(.1,.3)}]path picture bounding box.south west) to[bend left=50] ([shift={(-.1,.3)}]path picture bounding box.south east);\draw[black,-latex] ([shift={(0,.1)}]path picture bounding box.south) -- ([shift={(.3,-.1)}]path picture bounding box.north);}}}

    \begin{scope}[line width=0.5pt]

            \draw[red] (-4.5,1) -- (-3.5,1) node[midway,above] {$A$};
            \draw[red] (-4.5,1) -- (-5.5,2.5) node[left] {\textcolor{black}{$\Psi$}};

            \draw[blue] (-4.5,4) -- (-3.5,4) node[midway,above] {$G_0$}
             -- (-0.5,4);
            \draw[blue] (-4.5,4) -- (-5.5,2.5) node[midway,above] {};

			\draw[->] (-3,0) node[below] {$m$} -- (-3,0.5) ;

           \draw[] (-2.5,1) -- (-1.5,1)     node[midway,above] {$X_i$}; 
           
           \draw[->] (-1.5,1) --  (-1.5,2)     node[midway,above] {};
           \draw[->] (-1.5,1) -- (-0.5,1)     ; 

           \filldraw [] (-1.5,1) circle (1pt);

		\draw  (-3.5,1.5) rectangle (-2.5,0.5) node[midway] 
        {Enc};
		
		\draw  (-2,3) rectangle (-1,2) node[midway] { $P_{Y_r|X}$};

        \draw  (-0.5,1.5) rectangle (2,0.5) node[midway] { $P_{Y|XX_rY_r}$};

		\draw  (-0.5,4.5) rectangle (2,3.5) node[midway] { Relay Encoder} ;

        \draw[->] (-1,2.5) -| (0,2.5) node[right] {$Y_{r,i}$} -- (0,3.5) ;
        \draw[->] (0,2.5) -- (0,1.5) ;
        \draw[->] (1.5,3.5) -- (1.5,1.5) node[midway,right] {$X_{r,i}$};

        \filldraw [] (0,2.5) circle (1pt);

	    \draw[->] (2,1) -- (3,1) node[midway,above] {$Y_i$};

		\draw  (3,1.5) rectangle (4,0.5) node[midway] 
        {Dec};
        
        \draw[->] (4,1) -- (5,1) node[midway,above] {$\hat{m} $}; 

\end{scope}

\end{tikzpicture}
\caption{Zero-delay relay with entangled transmitters.}
\label{Figure:Causal_encoding}
\end{figure*}

We consider a model for the three-terminal network in 
Figure~\ref{Fig:relay_diag},
where Alice communicates with Bob through a relay. % 
Formally, a discrete memoryless unit-delay relay channel  $(\mathcal{X}\times\mathcal{X}_r,P_{YY_r|XX_r},
\mathcal{Y}\times\mathcal{Y}_r)$ is specified by finite
 input alphabets $\mathcal{X}$, $\mathcal{X}_r$ of the sender and the relay, respectively,
 output alphabets $\mathcal{Y}$, $\mathcal{Y}_r$ of the destination and the relay, respectively, and a collection of probability functions $P_{YY_r|XX_r}(\cdot,\cdot|x,x_r)$ on $\mathcal{Y}\times \mathcal{Y}_r$, for each input pair
 $(x,x_r)\in \mathcal{X}\times\mathcal{X}_r$.
Here, 
$X$, $X_r$, $Y$, and $Y_r$  are associated with the 
sender transmitter, the relay transmitter, the destination receiver, and the relay receiver, respectively.

The channel is assumed to be memoryless, i.e., given an input block
$(x^n,x_r^n)$, the channel output is distributed according to
\begin{align}
    P_{YY_r|XX_r}^n(y^n,y_r^n|x^n,x_r^n)\equiv 
    \prod_{i=1}^n P_{YY_r|XX_r}(y_i,y_{r,i}|x_i,x_{r,i})
    \,.
\end{align}
Our communication setting will also assume that the transmitters are provided with entanglement assistance. That is, the sender and the relay are provided with an entangled resource
$\Psi_{A G_0}$ a priori, before communication begins,
where $A$ is Alice's share, and $G_0$ is the relay's share.
The relay serves the purpose of assisting the transmission of information from Alice to Bob.

We provide the coding and capacity definitions for a classical relay channel, where the transmitters have access to pre-shared entanglement.

\subsection{Strictly-Causal Coding}
\label{Section:Coding_SC}
First, we consider communication via
a unit-delay relay channel $P_{YY_r|XX_r}$, where the relay can only access past received outputs, $Y_{r}^{i-1}$, as illustrated in 
 Figure~\ref{Figure:Strictly_causal_encoding}.

\begin{definition} 
\label{def:ClcapacityE}
A $(2^{nR},n)$  code for the unit-delay relay channel 
$P_{YY_r|XX_r}$ with entangled transmitters 
consists of the following: 
\begin{itemize}
\item 
A message set  $[2^{nR}]$, where $2^{nR}$ is % 
an integer, 

\item
a bipartite state $\Psi_{AG_0}$ of arbitrary dimension,

\item
a collection of encoding measurements $\{\mathsf{K}_{x^n|m}\}$ on 
$\mathcal{H}_A$,

\item
a sequence of strictly-causal relay encoding measurements 
$\{\mathsf{L}^{(i)}_{x_{r,i}|y_r^{i-1}}\}$, for 
$i\in [n]$,

\item	
a decoding function   $\text{Dec} :
\mathcal{Y}^n\to [2^{nR}]$. 
\end{itemize}
\end{definition}

The communication scheme is depicted in Figure~\ref{Figure:Strictly_causal_encoding}.
Alice selects a uniform message $m\in [2^{nR}]$ that is intended for the destination receiver, Bob.
Alice and the relay share $\Psi_{AG_0}$ a priori (the dimension depends on $n$ in general). 
She performs the measurement $\{\mathsf{K}_{x^n|m}\}$ on her share $A$ of the entanglement resource, 
and obtains a measurement outcome $x^n$.
At time $i$, Alice sends $x_i$.
The relay encodes $x_{r,i}$ in a strictly-causal manner. That is, having observed the past received outputs $y_{r,1},\ldots,y_{r,i-1}$, the relay performs the measurement $\{\mathsf{L}^{(i)}_{x_{r,i}|y_r^{i-1}}\}$ on the relay's entanglement share. The measurement maps
$G_{i-1}$ to $G_i$,
where $G_i$ is the residual quantum memory of the relay at time $i$, after the measurement.
The relay then transmits the measurement outcome $x_{r,i}$. This results in the joint distribution below:
\begin{align}
&p_{X^n X_{r}^n  Y^n Y_{r}^n|M}(x^n,x_{r}^n,y^n,y_{r}^n|m)=
\nonumber\\
&\trace\left[ \left(\mathsf{K}_{x^n|m}\otimes {{\mathsf{L}}}^{\text{s-c}}_{x_r^n|y_r^n}\right) \Psi_{A G_0} \left(\mathsf{K}_{x^n|m}\otimes {{\mathsf{L}}}^{\text{s-c}}_{x_r^n|y_r^n}\right)^{\dagger} \right]
 \nonumber\\
 &
\cdot P^n_{YY_r|XX_r}(y^n,y_{r}^n|x^n,x_{r}^n)
\label{Equation:Strictly_Causal_Distribution}
\intertext{% 
where}
&{\mathsf{L}}^{\text{s-c}}_{x_r^n|y_r^n}\equiv
\mathsf{L}^{(n)}_{x_{r,n}|y_{r}^{n-1}}\cdots\mathsf{L}^{(2)}_{x_{r,2}|y_{r,1}} \mathsf{L}^{(1)}_{x_{r,1}}
\,.
\end{align}
Let $p_{Y^n |M}(y^n|m)$ denote Bob's marginal distribution.
Then, the average error probability is given by
\begin{align}
    P_e^{(n)}\equiv \frac{1}{2^{nR}}\sum_{m=1}^{2^{nR}}\;
    \sum_{y^n:\;\text{Dec}(y^n)\neq m
    }
    p_{Y^n|M}(y^n|m)
    \,.
\end{align}

A $(2^{nR},n,\varepsilon)$  code % 
satisfies 
$
P_{e}^{(n)}% 
\leq\varepsilon $.  
A rate  $R> 0$ is % 
achievable   
if for every $\varepsilon,\delta>0$ and sufficiently large $n$, there exists a 
$(2^{n(R-\delta)},n,\varepsilon)$ code. 
 The  capacity of the unit-delay relay channel with entangled transmitters 
 $C^{(1)}_{\text{ET}}$ 
is defined as 
the supremum of achievable rates.

\begin{remark}
\label{Remark:Classical_Unit_Delay}
If we remove the entanglement resources, the setting reduces to the standard relay setting with strictly-causal encoding
\cite{vanderMeulen:71p,Sato1976,CoverElGamal:79p}. % 
\end{remark}

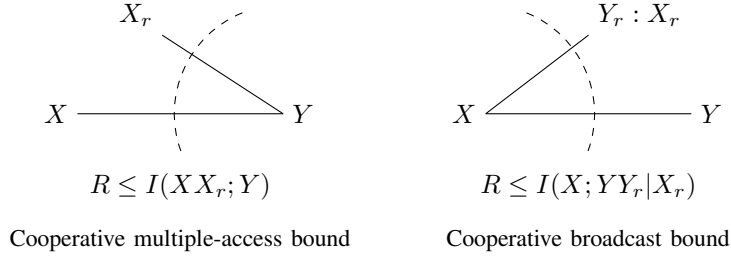
\begin{figure*}[htbp]
\centering
\begin{tikzpicture}[x=1cm, y=1cm]

  \begin{scope}[shift={(0,0)},scale=0.8]
    \coordinate (X1) at (0, 0);
    \coordinate (Y3) at (3.4, 0);
    \coordinate (X2) at (1.4, 1.3);

    \draw (X1) -- (Y3);
    \draw (X2) -- (Y3);

    \node[left] at (X1) {$X$};
    \node[right] at (Y3) {$Y$};
    \node[above left] at (X2) {$X_r$};

    \draw[dashed] ([shift={(112:1.8)}]Y3) arc (112:200:1.8);

    \node at (1.7, -1.2) {$R \leq I(X X_r; Y)$};
    \node at (1.7, -2.1) {\small{Cooperative multiple-access bound}};
  \end{scope}

  \begin{scope}[shift={(5.4,0)},scale=0.8]
    \coordinate (X1) at (0, 0);
    \coordinate (Y3) at (3.4, 0);
    \coordinate (Relay) at (1.7, 1.3);

    \draw (X1) -- (Y3);
    \draw (X1) -- (Relay);

    \node[left] at (X1) {$X$};
    \node[right] at (Y3) {$Y$};
    \node[above right] at (Relay) {$Y_r : X_r$};

    \draw[dashed] ([shift={(68:1.8)}]X1) arc (68:-20:1.8);

    \node at (1.7, -1.2) {$R \leq I(X; YY_r|X_r)$};
    \node at (1.7, -2.1) {\small{Cooperative broadcast bound}};
  \end{scope}

\end{tikzpicture}

\caption{Min-cut interpretation of the cutset bound.}
\label{Figure:Cutset_interpretation}
\end{figure*}

\subsection{Causal Coding}
\label{Section:Coding_Causal}
Now, we consider communication via
a zero-delay relay channel, where the relay can access both present and past received outputs, i.e., $Y_{r,1},\ldots,Y_{r,i}$, as in Figure~\ref{Figure:Causal_encoding}.  
In this case, the channel is specified by $P_{Y_r|X}\times P_{Y|XX_rY_r}$.

A $(2^{nR},n)$  code % 
is defined in a similar manner as in
Definition~\ref{def:ClcapacityE}.
Here, however, the relay's encoding is based on a sequence of causal relay encoding measurements 
$\breve{\mathsf{L}}^{(i)}_{x_{r,i}|y_r^{i}}$. 
At time $i$, % 
the relay receives $y_{r,i}$, and then encodes $x_{r,i}$ in a causal manner: Having observed the past and present received outputs $y_{r}^i$, the relay performs the measurement $\{\breve{\mathsf{L}}^{(i)}_{x_{r,i}|y_r^{i}}\}$, and then transmits the measurement outcome $x_{r,i}$. This results in a joint distribution
as in \eqref{Equation:Strictly_Causal_Distribution}, with $\mathsf{L}^{\text{s-c}}_{x_r^n|y_r^n}$ replaced by
\begin{align}
&\mathsf{L}^{\text{caus}}_{x_r^n|y_r^n}\equiv
\mathsf{L}^{(n)}_{x_{r,n}|y_{r}^{n}}\cdots\mathsf{L}^{(2)}_{x_{r,2}|y_{r,1} y_{r,2}} \mathsf{L}^{(1)}_{x_{r,1}|y_{r,1}} \,.
\end{align}
The error probability and achievable rates are defined in a similar manner, and
 the  capacity of the zero-delay relay channel with entangled transmitters 
 $C^{(0)}_{\text{ET}}$ 
is defined accordingly.

\begin{remark}
\label{Remark:Classical_Zero_Delay}
If we remove the entanglement resources, the setting reduces to the classical zero-delay relay setting
\cite{ElGamalHassanpour2005,ElGamalHassanpourMammen:07p}.
\end{remark}

We denote the capacity without assistance by $ C_{\text{unassisted}}^{(0)}$.

\section{Negative Results}
\label{Sec:Main_SC}

In this section, we focus on the unit-delay relay channel with entangled transmitters, where the relay only has access to past received observations. 
We show that the standard cutset upper bound holds for this channel.

\subsection{Cutset Bound}
\label{Subsection:Cutset}
The strictly-causal cutset rate is defined by \cite{CoverElGamal:79p} % 
\begin{align}
I_{\text{cut}}^{(1)}\equiv \max_{  p_{X X_r}   } 
\min\Big\{
I(X X_r;Y)
,\,
I(X;Y Y_r |X_r)
\Big\}
\label{eq:Cutset}
\end{align}
where the maximum is over the set of all input distributions $p_{X X_r}$, and the mutual informations are computed with respect to
$(X,X_r,Y,Y_r)\sim p_{XX_r}(x,x_r) P_{YY_r|XX_r}(y,y_r|x,x_r)$.
Cover and El Gamal \cite{CoverElGamal:79p} established that the cutset rate above is an upper bound on the capacity of the unit-delay relay channel in the classical setting, i.e., without entanglement resources (see \cite[Th. 16.1]{ElGamalKim:11b}).

\begin{remark}
\label{Remark:Cutset}
The cutset upper bound has a min-cut interpretation, as illustrated in Figure~\ref{Figure:Cutset_interpretation}. 
 A cut partitions the three terminals, namely, the sender, relay, and destination, into two subsets.
The multiple-access cut separates the sender and relay from the destination. It is named so because, across this cut, the sender and relay act as two cooperating transmitters communicating with a single receiver, the destination. This cut yields the bound
$
R \leq I(XX_r;Y)
$.
The broadcast cut separates the sender from the relay and destination, yielding
$
R \leq I(X;Y Y_r| X_r)
$.
\end{remark}

 Our first capacity result is given in the theorem below.
\begin{theorem}
\label{theo:Cutset}
The capacity  of the  unit-delay relay channel $P_{YY_r|XX_r}$ with entangled transmitters satisfies
\begin{align}
C_{\text{ET}}^{(1)}\leq
I_{\text{cut}}^{(1)} \,.
\end{align}
\end{theorem}
The proof is given in Subsection~\ref{Subsection:Cutset_Proof_SC} below.

\subsection{Consequences}
\label{subsection:cq_channel}
The cutset bound is tight for several channel classes, including:
degraded relay channels \cite[Th.~1]{CoverElGamal:79p}, reversely degraded relay channels \cite[Th.~2]{CoverElGamal:79p}, % 
semi-deterministic relay channels~\cite{ElGamalAref:82p,CoverKim:07,Kim:08}, and  relay channels with orthogonal-sender components~\cite{ElGamalZahedi:05p}. % 
Our results above imply that entanglement assistance between the sender and the relay does not improve capacity performance for none of these channel classes.

\subsection{Proof of Theorem~\ref{theo:Cutset}}
\label{Subsection:Cutset_Proof_SC}
Consider communication via
a classical unit-delay relay channel $P_{YY_r|XX_r}$, where % 
the transmitters are provided with % 
a pre-shared entanglement resource $\Psi_{A G_0}$. % 
We now describe the code in terms of completely-positive trace-preserving (CPTP) maps.
Let $\mathcal{P}_{XX_r\to YY_r}$ denote the CPTP map that represents the classical channel $P_{YY_r|XX_r}$ \cite[Sec. 4.6.4]{Wilde:17b}.

Suppose Alice prepares a classical message state,
\begin{align}
    \pi_{MM_0}=\frac{1}{2^{nR}} \sum_{m=1}^{2^{nR}} \ketbra{m}_M\otimes \ketbra{m}_{M_0}
    \,.
\end{align}
Hence, the initial state is 
\begin{align}
    \pi_{MM_0}\otimes \Psi_{AG_0}
    \,.
\end{align}
To encode the message,
she performs a measurement $\mathcal{K}_{M_0 A^n \to X^n}$, and transmits the measurement outcome $X^n$ over $n$ channel uses. 
This gives rise to the following c-q state,
\begin{align}
    \rho^{(0)}_{MG_0 X^n }= (\mathrm{id}_{MG_0}\otimes \mathcal{K}_{M_0 A^n \to X^n})(\pi_{MM_0}\otimes\Psi_{G_0 A^n}) \,.
\end{align}

The relay encodes in a strictly-causal manner.
At time $i$, the relay has access to the past received observations $Y_r^{i-1}$, and performs a measurement specified by a quantum instrument 
$\mathcal{L}^{(i)}$, hence 
\begin{align}
&
\rho_{M X_r^{i-1} Y_r^{i-1} G_{i} X_{r,i} X^n  Y^{i-1} }^{(i)}=
\nonumber\\
&
(\mathrm{id}_M\otimes 
\mathcal{L}^{(i)}_{X_r^{i-1} Y_{r,{i-1}} G_{i-1}   \rightarrow X_r^{i-1} Y_{r,{i-1}} G_i X_{r,i}}
\otimes
\mathrm{id}_{X^n  Y^{i-1}}
)
\nonumber\\
&
(\rho_{M X_r^{i-1} Y_{r,{i-1}} G_{i-1} X^n   Y^{i-1}   }^{(i-1)})
\,,
\end{align}
where $G_i$ is the residual quantum memory of the relay at time $i$, after the measurement.
The relay transmits $X_{r,i}$, and then receives
$Y_{r,i}$.

Since this is a c-q state, it can also be expressed as % 
\begin{align}
    &\rho_{M  Y^{i-1} Y_r^{i-1} X^n X_r^{i}  G_{i}   }^{(i)}=
    \nonumber\\
    &\sum_{m,y^{i-1},y_r^{i-1},x^{n},x_r^i}
    \Pr(m,y^{i-1},y_r^{i-1},x^{n},x_r^i)
    \nonumber\\
    &
    \ketbra{m,y^{i-1},y_r^{i-1},x^{n},x_r^i}
    \nonumber\\
    &
    \otimes \psi^{(m,y^{i-1},y_r^{i-1},x^{n},x_r^i)}_{G_i}
\end{align}
for some $\psi_{G_i}^{(\cdot)}$.
We can then append the channel outputs and write
\begin{align}
    &\rho_{M  Y^{i-1} Y_r^{i-1} X^n X_r^{i} Y_i Y_{r,i}  G_{i}   }^{(i)}=
    \nonumber\\
    &
    \sum_{m,y^{i-1},y_r^{i-1},x^{n},x_r^i}
    \Pr(m,y^{i-1},y_r^{i-1},x^{n},x_r^i)
    \nonumber\\
    &
    \ketbra{m,y^{i-1},y_r^{i-1},x^{n},x_r^i}
    \nonumber\\
    &
    \otimes
   \mathcal{P}_{XX_r\to YY_1}(x_i,x_{r,i})% 
    \otimes 
    \psi^{(m,y^{i-1},y_r^{i-1},x^{n},x_r^i)}_{G_i}
    \,.
\end{align}
This results in a final state $\rho_{M  Y^n Y_r^{n} X^n X_r^{n}    G_{n}   }^{(n)}$ at time $i=n$.

Based on Fano's inequality \cite[Sec. 2.10]{CoverThomas:06b}, % 
\begin{align}
    nR&\leq I(M;Y^n)% 
    +n\varepsilon_n \,.
    \label{Equation:Fano}
\end{align}
We note that this follows as in the standard setting,  because the decoder operates in exactly the same manner, as without entanglement. 

We now show two bounds: 
\begin{align}
I(M;Y^n)% 
&\leq 
\sum_{i=1}^n I(X_i X_{r,i}  ;Y_i)% 
\label{Equation:MAC_0}
\\
I(M;Y^n)% 
&\leq 
    \sum_{i=1}^n I(X_i;Y_i Y_{r,i}|X_{r,i})% 
    \label{Equation:Broadcast_0}
\end{align}
hereby referred to as the \emph{multiple-access} and \emph{broadcast} cuts, respectively (see Remark~\ref{Remark:Cutset}).

\subsubsection{Multiple-access cut}
By the chain rule,
\begin{align}
    I(M;Y^n)
    &=
    \sum_{i=1}^n I(M;Y_i|Y^{i-1})% 
    \nonumber\\
    &\leq
    \sum_{i=1}^n I( M Y^{i-1} X_i X_{r,i};Y_i)% 
    \nonumber\\
    &=
    \sum_{i=1}^n I(X_i X_{r,i}  ;Y_i)% 
\end{align}
where the last inequality holds since $(M,Y^{i-1})\Cbar (X_i, X_{r,i})\Cbar Y_i $ form a Markov chain. % 
Indeed, given $(X_i=x_i,X_{r,i}=x_{r,i})$, the channel output
$\mathcal{P}_{XX_r\to YY_r}(x_i,x_{r,i})$ has no correlation with 
$(M,Y^{i-1})$.
We have thus shown the multiple-access cut bound in \eqref{Equation:MAC_0}.

\subsubsection{Broadcast cut}
First, notice that 
\begin{align}
    I(M; Y^{n})% 
    \leq I(M;X_r^{n} Y_r^{n}   Y^{n})% 
    \,.
    \label{Equation:Broadcst_1}
\end{align}
To derive the broadcast-cut bound,
we introduce  two lemmas.
Define 
\begin{align}
    T_i&\equiv (G_i,X_r^{i-1}, Y_r^{i-1},   Y^{i-1})\\
    S_i&\equiv (T_i,X_{r,i})
\end{align}
for $i\in\{1,\ldots,n+1\}$, where $X_r^0$, $Y_r^0$, $Y^0$, $X_{r,n+1}$, and $G_{n+1}$ are degenerate (say, of dimension $1$). 
\begin{lemma}
\label{Lemma:I_S_Difference}
\begin{align}
    I(M;S_{i+1})_{\rho^{(i+1)}}-I(M;S_{i})_{\rho^{(i)}}\leq
    I(M;Y_i Y_{r,i}|S_i)_{\rho^{(i)}}
\end{align}
for $i\in \{1,\ldots,n\}$.
\end{lemma}

\begin{proof}[Proof of Lemma~\ref{Lemma:I_S_Difference}]
    At time $i+1$, Alice sends $X_{i+1}$, 
    and the relay performs a measurement
    on $(G_i,X_r^i,Y_r^i)$, producing $(G_{i+1},X_r^i,Y_r^i,X_{r,i+1})$.
    The channel appends the output $Y^{i+1} Y_r^{i+1}$.
    Therefore, $S_{i+1}% 
    $ is obtained from 
    $(S_i,Y_i,Y_{r,i})% 
    $ through the relay's measurement and the communication channel. 
    Then, by the data processing inequality \cite[Th. 11.9.4]{Wilde:17b},
    \begin{align}
        I(M;S_{i+1})_{\rho^{(i+1)}}&\leq I(M;S_{i} Y_i Y_{r,i})_{\rho^{(i)}}
        \nonumber\\
        &= I(M;S_{i} )_{\rho^{(i)}}+
         I(M; Y_i Y_{r,i}|S_{i})_{\rho^{(i)}}
    \end{align}
    from which the lemma immediately follows. 
\end{proof}

\begin{lemma}
\label{Lemma:Gi_Mutual_Information}
    \begin{align}
        I(M; Y_i Y_{r,i}| S_i)_{\rho^{(i)}}\leq
        I(X_i;Y_i Y_{r,i}|X_{r,i})% 
    \end{align}
\end{lemma}
\begin{proof}[Proof of Lemma~\ref{Lemma:Gi_Mutual_Information}]
    First, we add the channel input $X_i$ and apply the chain rule:
    \begin{align}
        &I(M; Y_i Y_{r,i}| T_i X_{r,i})_{\rho^{(i)}}
        \nonumber\\
        &\leq
        I(M X_i; Y_i Y_{r,i}| T_i X_{r,i} )_{\rho^{(i)}}
        \nonumber\\
        &=
        I( X_i; Y_i Y_{r,i}| T_i X_{r,i} )_{\rho^{(i)}}
        +
        I(M ; Y_i Y_{r,i}| X_i  X_{r,i} T_i)_{\rho^{(i)}} \,.
    \end{align}

The last information term is zero since the channel is memoryless. 
Specifically, given 
$(X_i=x_i,X_{r,i}=x_{r,i})$, the output state 
$\mathcal{P}_{XX_r\to YY_1}(x_i,x_{r,i})$ % 
has no correlation with 
$(M,T_i)% 
$.
Thus,
\begin{align}
        &I(M; Y_i Y_{r,i}| T_i X_{r,i})_{\rho^{(i)}}
        \nonumber\\
        &\leq
        I( X_i; Y_i Y_{r,i}|  X_{r,i} T_i)_{\rho^{(i)}}
        \nonumber\\
        &=
        H(Y_i Y_{r,i}|  X_{r,i} T_i)_{\rho^{(i)}}-H(Y_i Y_{r,i}| X_i X_{r,i} T_i)_{\rho^{(i)}}
        \,.
    \end{align}
    The first term is bounded by 
    $H(Y_i Y_{r,i}|  X_{r,i})$, based on entropy strong sub-additivity  \cite[Corollary 11.9.1]{Wilde:17b}. 
    Furthermore, since the channel is memoryless, the second term equals
    $H(Y_i Y_{r,i}| X_i X_{r,i} )$. Hence, 
\begin{align}
        &I(M; Y_i Y_{r,i}| S_i )_{\rho^{(i)}}
        \nonumber\\
        &\leq
        H(Y_i Y_{r,i}|  X_{r,i} )-H(Y_i Y_{r,i}| X_i X_{r,i} )
               \nonumber \\
        &=I(X_i; Y_i Y_{r,i}| X_{r,i})% 
    \end{align}
as we were set to prove.
\end{proof}
Together, the lemmas above imply 
    \begin{align}
    I(M;S_{i+1})-I(M;S_{i})\leq
    I(X_i;Y_i Y_{r,i}|X_{r,i})
    \,.
\end{align}
The sum over $i\in [n]$ is of telescopic form, which yields
\begin{align}% 
    I(M;S_{n+1})-I(M;S_{1})\leq
    \sum_{i=1}^n I(X_i;Y_i Y_{r,i}|X_{r,i})
    \,.
\end{align}% 
Now, notice that apart from degenerate systems,
$    S_{n+1}$ comprises $ (X_r^{n}, Y_r^{n},   Y^{n})$.
Furthermore, since at time $i=1$, the relay does not have any information, $I(M;S_{1})=I(M;G_1 X_{r,1})=0$. 
Hence,
\begin{align}
    I(M;X_r^{n}, Y_r^{n},   Y^{n})\leq
    \sum_{i=1}^n I(X_i;Y_i Y_{r,i}|X_{r,i}) \,.
    \label{Equation:Broadcst_2}
\end{align}
The broadcast-cut bound \eqref{Equation:Broadcast_0} now follows from \eqref{Equation:Broadcst_1} and \eqref{Equation:Broadcst_2}.

Together, \eqref{Equation:Fano}-\eqref{Equation:Broadcast_0} yield
\begin{align}
&R-\varepsilon_n
\nonumber\\
&
\leq 
\min\bigg\{\frac{1}{n}\sum_{i=1}^n I(X_i X_{r,i}  ;Y_i)
\,,
  \frac{1}{n}  \sum_{i=1}^n I(X_i;Y_i Y_{r,i}|X_{r,i})% 
  \bigg\}% 
\nonumber\\
&=\min\bigg\{ I(X_J X_{r,J}  ;Y_J|J)% 
\,,
 I(X_J;Y_J Y_{r,J}|X_{r,J} J)% 
  \bigg\}% 
\end{align}
where $J$ is uniformly distributed over 
$\{1,\ldots,n\}$. % 
We conclude the proof using the standard arguments:  
\begin{align}
    I(X_J X_{r,J}  ;Y_J|J)
    &=H(Y_J|J)-H(Y_J|X_J X_{r,J})
    \nonumber\\
    &\leq H(Y_J)-H(Y_J|X_J X_{r,J})
    \nonumber\\
    &=I(X_J X_{r,J}  ;Y_J) \,,
    \intertext{and similarly}
     I(X_J;Y_J Y_{r,J}|X_{r,J} J)
     &\leq I(X_J;Y_J Y_{r,J}|X_{r,J})
\end{align}
based on strong sub-additivity. 
The cutset bound follows as we identify $X$, $X_r$, $Y$, 
$Y_r$ as 
$X_J$, $X_{r,J}$, $Y_{J}$, $Y_{r,J}$, respectively.
\qed

\section{Positive Result}
\label{Sec:Main_Causal}
Here, we consider the zero-delay relay channel, where the relay  has access to both the present and past received observations. 
The channel is specified by a probability function of the form $P_{Y_r|X}\times P_{Y|XX_r Y_r}$.
We show that entanglement between the sender and relay's transmitters increases the capacity of this classical channel.

\begin{table}[t]
\centering
\renewcommand{\arraystretch}{1.2}
\setlength{\tabcolsep}{4pt}

\begin{minipage}[t]{0.4\columnwidth}
\centering
\caption{Classical Strategy% 
}
\label{tab:classical-magic-square}
\vspace{0pt}
{\arrayrulecolor{gray!80}
\begin{tabular}{|c|c|c|}
 \hline 
 $0$ & $0$ & $0$  \\  \hline
 $0$ & $0$ & $0$  \\  \hline
 $1$ & $1$ & $?$  \\  \hline
\end{tabular}
}
\end{minipage}
\hspace{-0.35cm}
\begin{minipage}[t]{0.58\columnwidth}
\centering
\caption{Quantum Strategy% 
}
\label{tab:quantum-magic-square}
\vspace{-1.6pt}
{\arrayrulecolor{gray!80}
\begin{tabular}{|c|c|c|}
\hline
 $\mathsf{X}\otimes \identity$
& $ \identity\otimes \mathsf{X}$
& $\mathsf{X}\otimes \mathsf{X}$
 \\ \hline
 $ \identity\otimes \mathsf{Z}$
& $\mathsf{Z}\otimes  \identity$
& $\mathsf{Z}\otimes \mathsf{Z}$
 \\ \hline
 $-\mathsf{X}\otimes \mathsf{Z}$
& $-\mathsf{Z}\otimes \mathsf{X}$
& $\mathsf{Y}\otimes \mathsf{Y}$
 \\
\hline
\end{tabular}
}
\end{minipage}
\end{table}

\subsection{Magic-Square Relay Channel}
The channel is constructed from the magic-square game, a pseudo-telepathy game in which quantum players can win with certainty, whereas classical players cannot \cite{brassard2005quantum}. The referee selects a cell $(r,c)$ uniformly at random from the nine cells of a $3\times 3$ square, sends the row index $r$ to Player~1, and sends the column index $c$ to Player~2. Then, Player~1 assigns three bits to row $r$, while Player 2 assigns three bits to column $c$. They win if the following conditions hold: their assignments agree at the intersection $(r,c)$, the bits assigned by Player~1 have even parity, and those assigned by Player~2 have odd parity. 
 Tables~\ref{tab:classical-magic-square} and \ref{tab:quantum-magic-square} illustrate classical and quantum strategies, respectively. 
No deterministic classical strategy can satisfy these conditions for every pair $(r,c)$. % 
In contrast, the following quantum strategy wins with probability (w.p.) $1$:
\begin{itemize}
\item
Before the game begins, the players share two maximally entangled qubit pairs:
\begin{align}
\ket{\phi_{A'A''G'G''}}
&=
\ket{\Phi^+}_{A'G'}
\otimes
\ket{\Phi^+}_{A''G''}
\,,
\label{eq:Magic_A1A2}
\end{align}
where Player~1 and Player~2 hold $A'A''$ and   $G'G''$, respectively.

\item
Upon receiving $r$, Player~1 jointly measures the three observables in row $r$ of Table~\ref{tab:quantum-magic-square}. Similarly, upon receiving  $c$, Player~2  measures the  observables in column $c$. These joint measurements are well defined because the three observables within each row and within each column commute. Each measurement outcome is converted into a bit according to
$
+1\mapsto 0
$ and 
$
(-1)\mapsto 1
$.
\end{itemize}
For example, if $(r,c)=(1,3)$, Player~1 measures
$\mathsf{X}\otimes\identity$,
$\identity\otimes\mathsf{X}$, and
$\mathsf{X}\otimes\mathsf{X}$, whereas Player~2 measures
$\mathsf{X}\otimes\mathsf{X}$,
$\mathsf{Z}\otimes\mathsf{Z}$, and
$\mathsf{Y}\otimes\mathsf{Y}$.
The product of their observables
are $+\identity$ and $-\identity$, respectively.
Consequently, Player~1's output has even parity and Player~2's output has odd parity. 
The % 
state in~\eqref{eq:Magic_A1A2} % 
guarantees that their outcomes are identical % 
at the intersection $(r,c)$. Hence, this strategy wins the magic-square game w.p. $1$.

In a general nonlocal game, a referee draws a pair of questions
$
(q_0,q_1)\in\mathcal{Q}_0\times\mathcal{Q}_1
$
 at random, and sends the respective question to each player.  The players produce their answers $a_0\in\mathcal{A}_0$ and $a_1\in\mathcal{A}_1$, without communicating. 
The players win if
\[
(q_0,q_1,a_0,a_1)\in\mathscr{G},
\]
where
$\mathscr{G}\subseteq
\mathcal{Q}_0\times\mathcal{Q}_1\times
\mathcal{A}_0\times\mathcal{A}_1$
denotes the winning set. The CHSH and magic-square games are particular instances of this framework \cite{scarani2019bell}.
Leditzky et al.~\cite{LeditzkyAlhejjiLevinSmith:20p} constructed a multiple-access channel whose output is noiseless when the players' answers satisfy the winning condition, and completely noisy otherwise. We use a similar idea.

We define a classical relay channel
$P_{Y_r|X}P_{Y|XX_rY_r}$ with alphabets
\begin{subequations}
    \label{Equation:Magic_Square_Relay}
\begin{align}
\mathcal{X}
&=\mathcal{Q}_0\times\mathcal{A}_0\,,\\
\mathcal{Y}_r
&=\mathcal{Q}_1 \,,\;
\mathcal{X}_r
=\mathcal{A}_1\,,\\
\mathcal{Y}
&=\mathcal{Q}_0\,.
\end{align}
The relay observation is drawn uniformly and independently of the channel input:
\begin{align}
P_{Y_r|X}
(q_1|(q_0,a_0))
=
\frac{1}{|\mathcal{Q}_1|}.
\label{eq:relay-question}
\end{align}
The destination channel is specified by
\begin{multline}
P_{Y|XX_rY_r}
(
\hat q_0
|
(q_0,a_0),a_1,q_1
)
\\
=
\begin{cases}
\delta_{\hat q_0,q_0},
& (q_0,q_1,a_0,a_1)\in\mathscr{G},\\[1mm]
\dfrac{1}{|\mathcal{Q}_0|},
& (q_0,q_1,a_0,a_1)\notin\mathscr{G}.
\end{cases}
\label{Equation:MS_channel}
\end{multline}
\end{subequations}
Thus, the source input $X=(q_0,a_0)$, relay observation $Y_r=q_1$, and relay input $X_r=a_1$ jointly specify a play of the game. If this play satisfies the winning condition, then the destination recovers $q_0$ perfectly, namely, $Y=q_0$ with probability one. Otherwise, $Y$ is uniformly distributed over $\mathcal{Q}_0$ and therefore carries no information about $q_0$.

The magic-square game is specified by questions $q_0,q_1$ from 
$\mathcal{Q}_0=\mathcal{Q}_1=\{1,2,3\}$, answers $a_0$, $a_1$ from 
$\mathcal{A}_0=\mathcal{A}_1=\{0,1\}^3$, 
and the winning set 
$
\mathscr{G}
=
\{
(q_0,q_1,(a_0[j],a_1[j])_{j=1,2,3})
:
a_0[q_1]=a_1[q_0]$,
 $
\sum_{j=1}^3 a_0[j]=0 \mod 2
$, and $
\sum_{j=1}^3 a_1[j]=1 \mod 2
\}
$.

\begin{remark}
In the game-based MAC of Leditzky et al.~\cite{LeditzkyAlhejjiLevinSmith:20p}, the two senders transmit the question-answer pairs $(Q_0,A_0)$ and $(Q_1,A_1)$. Thus, their joint channel input specifies the entire play. In our relay channel, the relay observes $Q_1$ and transmits only $A_1$. Its causal encoder therefore implements the second player's response strategy.
\end{remark}

We show an entanglement-assisted gain in the communication rate of a single sender through a relay.

\begin{theorem}[Entanglement advantage]
    Consider the zero-delay relay channel $P_{Y_r|X}\times P_{Y|XX_r Y_r}$, as specified in 
    \eqref{Equation:Magic_Square_Relay}.
    Then, the capacity without assistance satisfies
    \begin{align}
        C_{\text{unassisted}}^{(0)}&\leq \frac{8}{9} \log 3
        \intertext{and the capacity with entangled transmitters is }
        C_{\text{ET}}^{(0)}&= \log 3
        \,.
    \end{align}
\end{theorem}
The proof is given in Subsections~\ref{Subsection:Classical_Bound}-\ref{Subsection:ET_Capacity} below.

\subsection{Classical Bound}
\label{Subsection:Classical_Bound}
First, we analyze the classical setting, % 
without quantum resources. 
According to the causal cutset bound \cite{ElGamalHassanpour2005},
\textcolor{red}{}
\begin{align}
    C_{\text{unassisted}}^{(0)}\leq 
\max_{  p_{UX} p_{ X_r|UY_r}   } 
\min\Big\{
I(UX;Y)
,\,
I(X; YY_r|U)
\Big\}
\label{eq:Cutset_causal_def}
\end{align}% 
where the maximum is over the set of all input distributions $p_{UX} p_{ X_r|UY_r}$ (see \cite[Th. 2]{ElGamalHassanpourMammen:07p}). Here, $U$ is an auxiliary variable, $\abs{\mathcal{U}}\leq \abs{\mathcal{X}} \abs{\mathcal{X}_r}+1$, and the mutual informations are computed with respect to
$(U,X,Y_r,X_r,Y)\sim
p_{UX}(u,x) P_{Y_r|X}(y_r|x)\cdot
p_{X_r|UY_r}(x_r|u,y_r) P_{Y|XX_rY_r}(y|x,x_r,y_r)$.

\subsubsection{Broadcast-cut bound}
We now focus on the broadcast cut bound: 
\begin{align}
    C_{\text{unassisted}}^{(0)}&\leq  
\max_{p_U  p_{X|U} p_{ X_r|UY_r}   } 
I(X; YY_r|U)
\\
&= 
\max_{  p_{X} p_{ X_r|Y_r}   } 
I(X; YY_r)
\label{eq:Broadcast_Cut_causal}
\end{align}% 
where we can remove the time-sharing variable $U$ since $I(X;YY_r|U)=\sum_u p_U(u)I(X;YY_r|U=u)$.

For the magic-square channel, 
\begin{multline}
 (X,Y_r,Y) \sim   p_{X}((q_0,a_0)) \cdot \frac{1}{3}
   \\ \cdot
p_{X_r|Y_r}(a_1|q_1) P_{Y|XX_rY_r}(\hat{q}_0|(q_0,a_0),q_1,a_1) \,.
\end{multline}
Since $Y_r$ is statistically independent of $X$,
\begin{align}
    C_{\text{unassisted}}^{(0)}
    &\leq 
\max_{  p_{X} p_{ X_r|Y_r}   } 
I(X; Y|Y_r) 
\label{eq:Cutset_causal_1}
\end{align}
as $I(X;Y_r)=0$.
Denote the broadcast-cut upper bound by 
\begin{align}
    I_{\text{broadcast}}^{(0)}\equiv     \max_{p_{Q_0}}
\max_{  p_{A_0|Q_0} p_{ A_1|Q_1}   } 
I(Q_0 A_0  ; \widehat{Q}_0 |Q_1) \,.
\end{align}

    Our analysis is inspired by the approach of Seshadri et al. \cite{SeshadriLeditzkySiddhuSmith}, who study multi-user channels constructed from nonlocal games and derive sum-rate bounds by relating the induced channel noise to limitations on the winning probability of the underlying game. The setting considered here is different in two respects. First, we analyze the broadcast cut of a zero-delay relay channel, which leads to the asymmetric optimization above,
rather than a multiple-access sum-rate expression.  Second, instead of relaxing the problem using only the maximum classical winning probability of the magic-square game, we retain the structure of the classical response strategies and exploit the parity constraints of the magic square. This yields a relay-specific bound that is tighter than the bound obtained from the winning probability alone.

\subsubsection{Constrained Optimization}
Define the variables
\begin{align}
    &\pi_j=\Pr(Q_0=j) 
\intertext{and}
    & w_{j,k}=\Pr(\text{win}|Q_0=j,Q_1=k)
\end{align}
for $j,k\in \{1,2,3\}$. 
Then, let $\bm{\pi}=(\pi_j)_{j\in\{1,2,3\}}$ and
$\bm{w}_k=(w_{j,k})_{j\in\{1,2,3\}}$ denote the corresponding column vectors, for $k\in\{1,2,3\}$.

For every classical strategy over a unifom table, $\bm{\pi}=\frac{\bm{1}}{3}$,
\begin{align}
    \frac{8}{9}&\geq\Pr(\text{win})
    = \frac{1}{9}\sum_{j,k} w_{j,k} \,.
\end{align}
This yields the following optimization constraint:
\begin{align}
    \sum_{j,k} w_{j,k}\leq 8 \,.
\end{align}

Now, let $k$ be a fixed column of the $3\times 3$ square (see Table~\ref{tab:classical-magic-square}).
Given $Q_1=k$, the winning probability is 
\begin{align}
    p_{\text{W},k}&\equiv \Pr(\text{win}|Q_1=k)
    \nonumber\\
    &=\bm{\pi}^T \bm{w}_{k} \,,
    \intertext{and the losing probability is }
    \lambda_k &% 
    \equiv 1-\bm{\pi}^T \bm{w}_{k}
    \,.
\end{align}
Bob's output entropy is then
\begin{align}
    & H(\widehat{Q}_0|Q_1=k)
    =H\left( \left\{\pi_\ell w_{\ell,k}+\frac{1-\bm{\pi}^T \bm{w}_k}{3} \right\} \right) \,.
\end{align}
Whereas, when conditioning on the entire play,
\begin{align}
    H(\widehat{Q}_0|A_0A_1 Q_0,Q_1=k)
    &=\lambda_{k} \log 3
    \,.
\end{align}
Thus,
\begin{align}
&I(A_0 Q_0   ; \widehat{Q}_0 |Q_1=k)\leq 
    I(A_0A_1 Q_0;\widehat{Q}_0|Q_1=k)
     \nonumber\\
    &=
    H\left( \left\{\pi_\ell w_{\ell,k}+\frac{1-\bm{\pi}^T \bm{w}_k}{3} \right\} \right)
    -\lambda_k \log 3 \,.
\end{align}
We would like to maximize $I(A_0 A_1 Q_0   ; \widehat{Q}_0 |Q_1)$, i.e., the average value of $I(A_0 A_1 Q_0   ; \widehat{Q}_0 |Q_1=k)$ over $k\in\{1,2,3\}$.

\subsubsection{Convex Relaxation}
Define a function 
$f:\mathscr{P}(\mathcal{Q}_0)\times [0,1]^{3}\to \mathbb{R} $ by
\begin{align}
    f(\bm{\pi},\bm{v}) 
        &=H\left( \left\{\pi_\ell v_{\ell}+\frac{1-\bm{\pi}^T \bm{v}}{3} \right\} \right)
    -[1-\bm{\pi}^T \bm{v}] \log 3
    \,.
\end{align}
Since $f$ is concave, we have by Jensen's inequality that
\begin{align}
    I(A_0A_1 Q_0;\widehat{Q}_0|Q_1)&=\frac{1}{3}\sum_{k=1}^3 f(\bm{\pi},\bm{w}_k)
    \nonumber\\
    &\leq  
    f(\bm{\pi},\bm{v} )
\end{align}
for
\begin{align}
    \bm{v}&=\frac{1}{3}\sum_{k=1}^3\bm{w}_k \,.
\end{align}

We deduce that the broadcast cut bound $I_{\text{broadcast}}^{(0)}$ is bounded by 
$f^*$, where $f^*$ is the optimal value of the following optimization problem: 
\begin{subequations}
\begin{align}
    \max_{\bm{\pi},\bm{v}} \quad & f(\bm{\pi},\bm{v}) 
    \nonumber
    \\
\textrm{s.t.} \quad 
& 
\pi_j\geq 0 \,,\;
\sum_{j=1}^3 \pi_j=1
\\
&v_{j}\geq0    
\,,\;
\sum_{j=1}^3  v_{j}\leq \frac{8}{3} \,.
\end{align}
\end{subequations}

Observe that the objective function can also be expressed as
\begin{align}
    &f(\bm{\pi},\bm{v})=g(\bm{t})
    \intertext{for}
    &t_j=\pi_j v_j \,,\; j\in\{1,2,3\} \,,
\end{align}
where the function 
$g: [0,1]^{3} \to\mathbb{R}$ is defined by
\begin{align}
    g(\bm{t}) 
    &=
    H  \left(\left\{ t_\ell+\frac{1-\bm{1}^T \bm{t}}{3} \right\}\right)
    -[1-\bm{1}^T \bm{t}] \log 3
    \,.
\end{align}

Next, consider the optimization constraint. By the Cauchy–Schwarz inequality,
\begin{align}
   \left( \sum_{j=1}^3 \sqrt{t_j} \right)^2
   &= \left( \sum_{j=1}^3 \sqrt{\pi_j} \sqrt{v_j} \right)^2
   \nonumber\\
   &\leq \left( \sum_{j=1}^3 \pi_j  \right)\left( \sum_{j=1}^3 v_j  \right)
      \nonumber\\
   &\leq  \frac{8}{3} \,.
\end{align}
Thus, $f^*$ is upper bounded by % 
\begin{subequations}
\begin{align}
    \max_{\bm{t}} \quad & g(\bm{t}) \nonumber\\
\textrm{s.t.} \quad 
& 
t_j\geq 0 \,,\;
\sum_{j=1}^3 t_j\leq 1    
\\
&\left( \sum_{j=1}^3 \sqrt{t_j} \right)^2\leq \frac{8}{3} \,.
\label{Equation:t_sqrt_constaint}
\end{align}
\end{subequations}

\begin{figure*}[htbp]
\centering
\begin{minipage}{0.45\textwidth}
  \centering
  \includegraphics[height=0.6\linewidth]{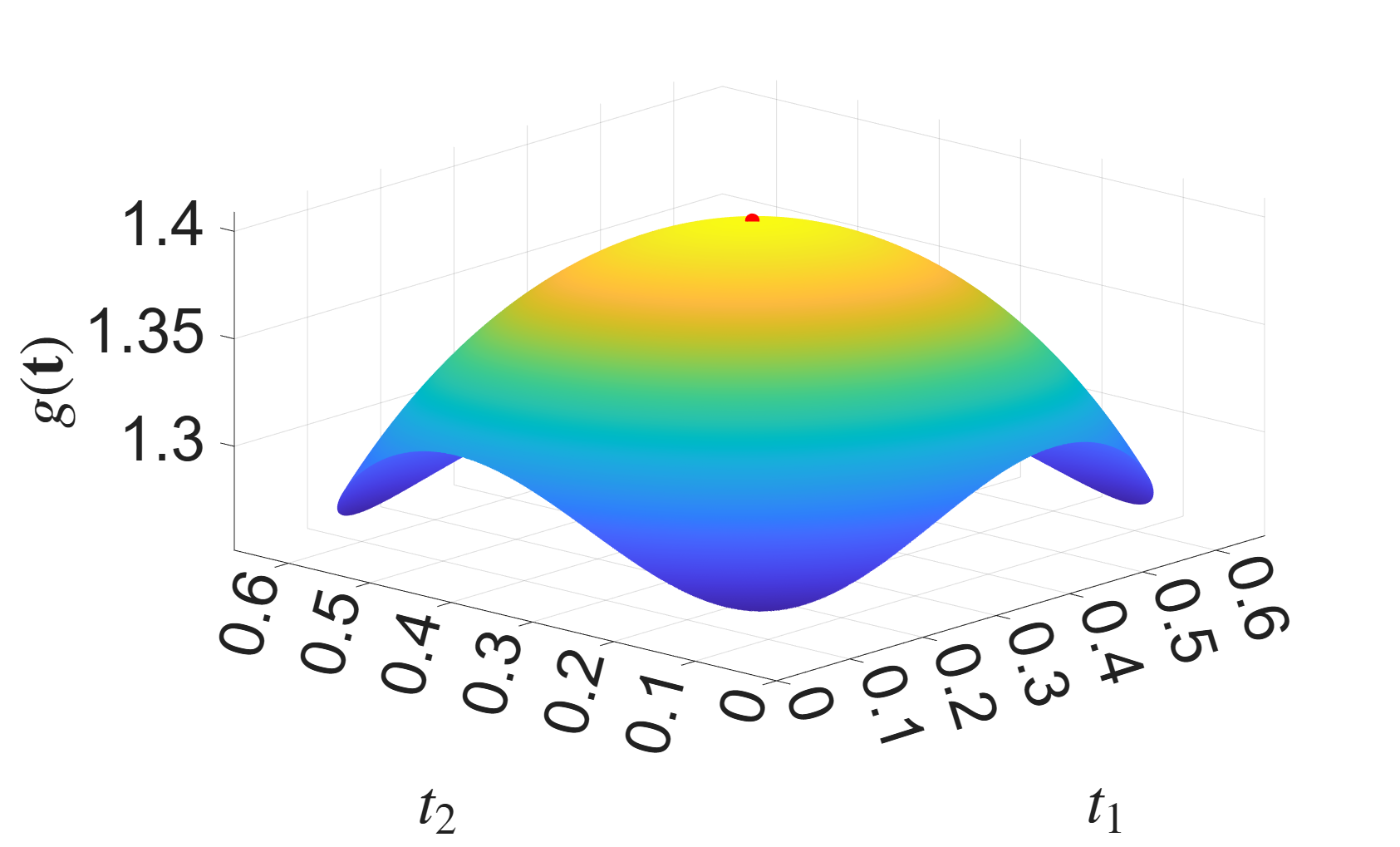}
  \\
  (a) Surface 
\end{minipage}
\hspace{-1cm}
\begin{minipage}{0.45\textwidth}
  \centering
  \includegraphics[height=0.6\linewidth]{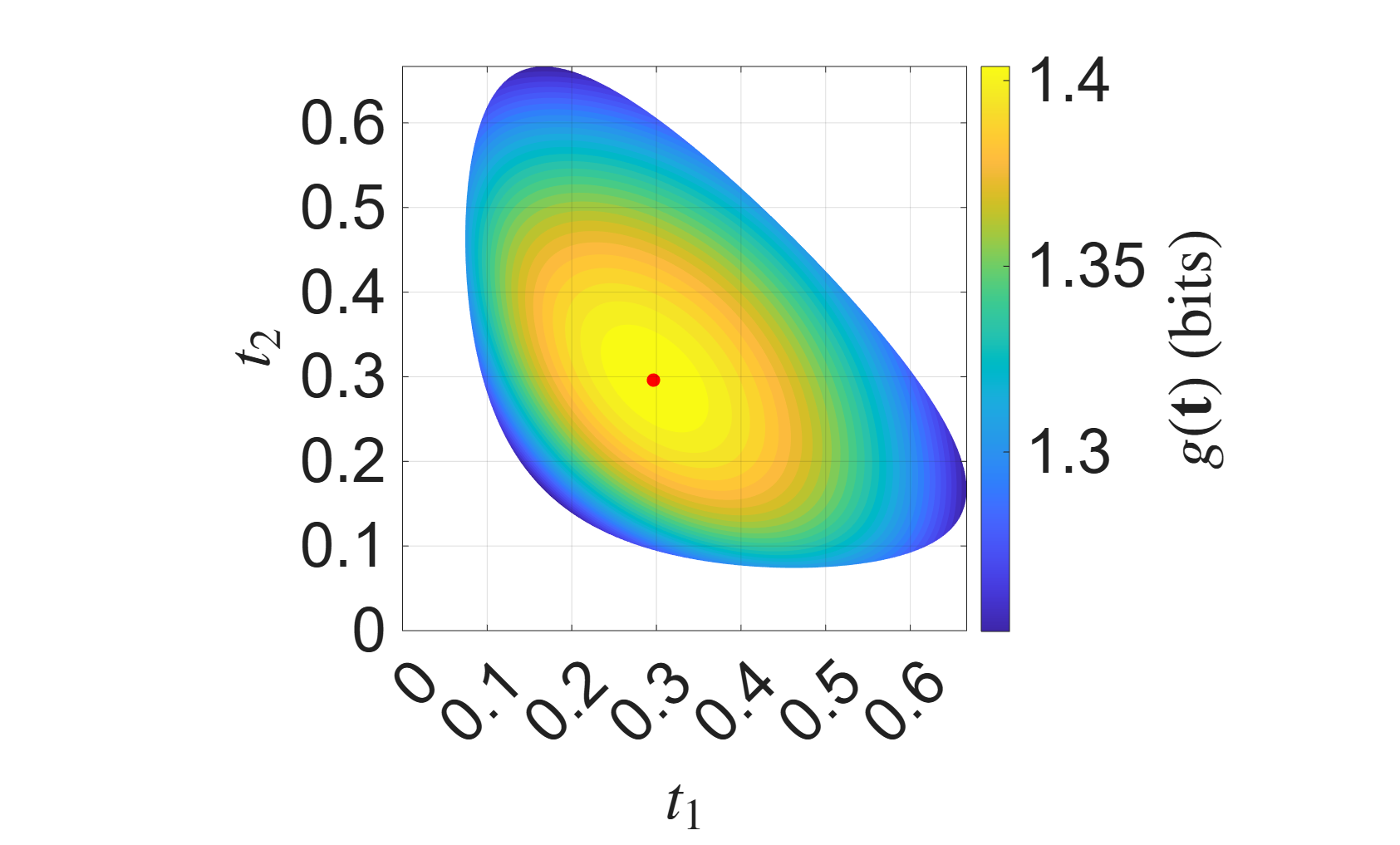}
  \\
  (b) Heatmap
\end{minipage}
\caption{The objective function $g(\mathbf{t})$ over the feasible set  in \eqref{Equation:reduced_optimization_g}. The red point indicates the maximum, $g\left(\frac{8}{27},\frac{8}{27},\frac{8}{27}\right)=\frac{8}{9}\log 3$.}
\label{Figure:Objective}
\end{figure*}

\subsubsection{Equality constraint}
We now claim that the last constraint, \eqref{Equation:t_sqrt_constaint}, can be changed to equality: 
\begin{align}
    &\left( \sum_{j=1}^3 \sqrt{t_j} \right)^2= \frac{8}{3}
    \,.
\label{Equation:t_sqrt_equality}
\end{align}
Consider a feasible $\bm{t}$.
We separate between two cases. 
First, assume that
\begin{align}
    \sum_{j=1}^3 t_j<1
    \,.
\end{align}
Suppose that we replace every $t_j$ by 
\begin{align}
    t_j^+=t_j+\delta \,.
\end{align}
This results in an increase of $3\delta$ to the sum: $\sum_j t_j^+=\sum_j t_j+3\delta$.
Observe that the entropy argument remains unchanged, as
\begin{align}
    t_\ell+\delta+\frac{1-(\bm{1}^T \bm{t}+3\delta)}{3}
    &=t_\ell+\frac{1-\bm{1}^T \bm{t}}{3}
    \,.
\end{align}
Thus,
\begin{align}
    g(\bm{t}^+) 
    &=g(\bm{t})+3\delta\log 3 
    \nonumber\\
    &>g(\bm{t}) \,.
\end{align}
Then, we can increase $\delta$ until either 
\eqref{Equation:t_sqrt_equality} holds, or 
$\sum\limits_{j=1}^3 t_j^+=1$. 

Next, consider the case of
\begin{align}
    \sum_{j=1}^3 t_j=1 \,.
\end{align}
Replace every $t_j$ by 
\begin{align}
    t_{\lambda,j}=(1-\lambda)t_j+\lambda\frac{1}{3} \,.
\end{align}
Observe that the first constraint still holds, as
\begin{align}
    \sum_{j=1}^3 t_{\lambda,j} &=(1-\lambda)\sum_{j=1}^3 t_j+3\cdot\lambda\frac{1}{3}
    =1 \,.
\end{align}
That is, $\bm{1}^T \bm{t}_\lambda=1$. 
Thus, by entropy concavity,
\begin{align}
    g(\bm{t}_\lambda) 
    &=H  (\bm{t}_\lambda)
    \nonumber\\
    &\geq 
    (1-\lambda) H  ( {\bm{t}} )+
    \lambda \log 3
    \nonumber\\
    &\geq  H  ( {\bm{t}} )
    \nonumber\\
    &=  g  ( {\bm{t}} ) \,.
\end{align}

As for the last constraint, by sqare-root concavity,
\begin{align}
     \sum_{j=1}^3 \sqrt{{t}_{\lambda,j}} 
     &\geq
     \sum_{j=1}^3 \left( (1-\lambda) \sqrt{ t_j}+\lambda\sqrt{\frac{1}{3}} \right) 
      \nonumber\\
     &=
     (1-\lambda)\sum_{j=1}^3 \sqrt{{t}_j} 
     +\lambda\cdot  \sqrt{3}
     \nonumber\\
     &>
     \sum_{j=1}^3 \sqrt{{t}_j} 
\label{Equation:t_sqrt_inequality}
\end{align}
by \eqref{Equation:t_sqrt_constaint}.
As $\lambda$ increases, $\sum_{j=1}^3 \sqrt{{t}_{\lambda,j}} $ continuously increases towards $\sqrt{3}$.
Thus, we can choose $\lambda$ such that the equality constraint \eqref{Equation:t_sqrt_equality} holds for $\bm{t}=\bm{t}_\lambda$.

Our optimization problem now reduces to the following: 
\begin{subequations}
\begin{align}
    \max_{\bm{t}} \quad & g(\bm{t}) \nonumber\\
\textrm{s.t.} \quad 
& 
t_j\geq 0 \,,\;
\sum_{j=1}^3 t_j\leq 1    
\\
&\left( \sum_{j=1}^3 \sqrt{t_j} \right)^2= \frac{8}{3}
\label{Equation:t_sqrt_reduced}
\end{align}
\label{Equation:reduced_optimization_g}
\end{subequations}
Figure~\ref{Figure:Objective} depicts the objective function over the feasible set, where the red dot indicates the maximal value.
One may verify numerically that the solution is uniform, namely, $t_1=t_2=t_3=\frac{8}{27}$, and the
maximal value is 
$g^*=\frac{8}{9}\log 3$.

We will now show this analytically.
Note that $g(\bm{t})$ can be expressed as
\begin{align}
g(\bm{t}) 
    &=(\bm{1}^T \bm{t}) \log 3-\left[\log 3- H  \left( \left\{ t_\ell+\frac{1-\bm{1}^T \bm{t}}{3} \right\} \right)
    \right] \,.
\end{align}

\subsubsection{Quadratic entropy bound}
We derive a general lower bound on $\log 3-H(\bm{p})$.
\begin{proposition}[Quadratic entropy bound]
\label{Lemma:Quadratic_Entropy_Bound}
    For every probability vector $\bm{p}\in\mathbb{P}_3$, 
    \begin{align}
      \log3-H(\bm{p})
    \geq 2\left( \sum_{j=1}^3 p_j^2-\frac{1}{3} \right)  \,, 
    \end{align}
    with equality for $\bm{p}=\left(\frac{1}{3},\frac{1}{3},\frac{1}{3} \right)$,
    $\left(\frac{1}{6},\frac{1}{6},\frac{2}{3} \right)$, and their permutations. 
\end{proposition}

The proposition can be obtained as a consequence of index of coincidence inequalities \cite{HarremoesTopsoe:01p}.
A simpler proof is given below. 
Define a function $\Delta:\mathbb{P}_3\to \mathbb{R}$ by 
\begin{align}
    \Delta(\bm{p})=\sum_{j=1}^3 p_j \log(3p_j)
    -2\left( \sum_{j=1}^3 p_j^2-\frac{1}{3} \right)
    \,.
\end{align}
We would like to find all probability vectors $\bm{p}$ that minimize $\Delta$. 
We note that $\Delta$ is continuous and the simplex is compact, hence a global minimizer exists.

We begin with the following lemma.
\begin{lemma}
\label{Lemma:Distinct_Coordinates}
    If $\bf{p}$ minimizes $\Delta$, then % 
    the following holds:
    \begin{enumerate}[1) ]
        \item 
        there exists a constant $c\in\mathbb{R}$ such that
        \begin{align}
\log(p_j)-4p_j=c
\label{Equation:KKT_Consequence_Lemma}
\end{align}
for $j\in\{1,2,3\}$.

\item
The vector $(p_1,p_2,p_3)$
consists of at most two distinct coordinates. 
    In other words, there exist two indices  $j$ and $\ell$, such that $p_j=p_{\ell}$ and $j\neq \ell$. 
    \end{enumerate}
    
\end{lemma}

The proof of Lemma~\ref{Lemma:Distinct_Coordinates} is given in Appendix~\ref{Appendix:Distinct_Coordinates}, based on first-order necessary conditions for optimality. 

As a consequence of part 2, a minimizer must have the form % 
\begin{align}
    \bm{p}=(\alpha,\alpha,1-2\alpha)
\end{align}
for some $\alpha\in \left(0,\frac{1}{2}\right)$, up to a permutation.
Then, by \eqref{Equation:KKT_Consequence_Lemma}, 
\begin{align}
    \log\alpha-4\alpha=
    \log(1-2\alpha)-4(1-2\alpha)
\end{align}
hence,
\begin{align}    \log\left(\frac{1-2\alpha}{\alpha}\right)&=
4(1-3\alpha) \,.
\label{Equation:Log_condition}
\end{align}
We observe that equality holds for 
$\alpha=\frac{1}{3}$, $\frac{1}{4}$ and $\frac{1}{6}$. 
\begin{lemma}
    \label{Lemma:Solution_Number}
    Equation~\eqref{Equation:Log_condition} has at most three solutions. 
\end{lemma}

\begin{proof}[Proof of Lemma~\ref{Lemma:Solution_Number}]
The proof follows by invoking Rolle's theorem.
Consider the function
\begin{align}
     \xi(x)=   \log\left(\frac{1-2x}{x}\right)-
4(1-3x)
\end{align}
on $\left(0,\frac{1}{2}\right)$.
Its first derivative $\xi'(x)$ % 
vanishes when 
\begin{align}
   2x^2-x+\frac{1}{12\ln 2}=0 \,.
   \label{Equation:Quadratic}
\end{align}
Since this is a quadratic equation, $\xi'$ has at most two zeros.

We would like to show that there are at most three values for which $\xi(x)=0$.
Assume by contradiction that there exist % 
\begin{align}
    &x_1<x_2<x_3<x_4
\intertext{such that}
&\xi(x_1)=\xi(x_2)=\xi(x_3)=\xi(x_4)=0 \,.
\end{align} 

By Rolle's theorem, 
there exists $c_i\in (x_i,x_{i+1})$ 
such that 
$\xi'(c_i)=0$, for $i\in \{1,2,3\}$.
Yet, there cannot be three different solutions to the quadratic equation \eqref{Equation:Quadratic}. 
Based on this contradiction, we deduce that our earlier assumption is false, and there cannot be more than three solutions.  
\end{proof}

As an immediate consequence of the lemma above, $\Delta$ has the following
stationary points, 
\begin{align}
    \bm{p}=\left(\frac{1}{3},\frac{1}{3},\frac{1}{3} \right)
    \,,\;
    \left(\frac{1}{6},\frac{1}{6},\frac{2}{3} \right)
    \,,\;
    \left(\frac{1}{4},\frac{1}{4},\frac{1}{2} \right) \,,
\end{align}
up to permutations, yielding the values
$0$, $0$, and $\log 3-\frac{19}{12}>0$, respectively. 
Hence, the minimal value is
\begin{align}
    \Delta_{\min}=0 \,.
\end{align}
Notice that $\Delta(\bm{p})$ can also be expressed as 
\begin{align}
    \Delta(\bm{p})
    &=\log3-H(\bm{p})
    -2\left( \sum_{j=1}^3 p_j^2-\frac{1}{3} \right)
    \,.
\end{align}
Proposition~\ref{Lemma:Quadratic_Entropy_Bound} immediately follows.

Based on the quadratic entropy bound in Proposition~\ref{Lemma:Quadratic_Entropy_Bound}, 
 our objective function 
$g(\bm{t})$ is bounded by
\begin{align}
    g(\bm{t})&\leq
    (\bm{1}^T\bm{t}) \log 3-2\left[\sum_{j=1}^3  \left(  t_j+\frac{1-\bm{1}^T\bm{t}}{3}  \right)^2-\frac{1}{3}
    \right] 
    \nonumber\\
   &= 3(\mu \log 3-2\sigma^2)
  \label{Equation:Objective_Bound_mu_sigma}  \intertext{where we define the ``empirical moments":}
    \mu&\equiv \frac{1}{3} \sum_{j=1}^3 t_j= \frac{\bm{1}^T\bm{t}}{3}
    \,,
    \\
    \sigma^2 &\equiv \frac{1}{3}\left[\sum_{j=1}^3  \left(  t_j-\mu+\frac{1}{3}  \right)^2-\frac{1}{3}
    \right] 
    \,.
\end{align}

\subsubsection{Change of variables}
As we set
\begin{align}
    q_j=\sqrt{t_j}
    \label{Equation:y_j}
\end{align}
we have 
\begin{align}
    \mu&=\frac{1}{3}\sum_{j=1}^3 q_j^2 \,,
    \label{Equation:mu_y}
    \\
    \sigma^2 &\equiv \frac{1}{3}\left[\sum_{j=1}^3  \left(  q_j^2-\mu+\frac{1}{3}  \right)^2-\frac{1}{3}
    \right] 
    \,,
    \label{Equation:sigma2_y}
\end{align}
and the constraints become
\begin{subequations}
\begin{align}
& 
q_j\geq 0 \,,\;
\mu\leq \frac{1}{3}    
\label{Equation:y_mu_Constraint}
\\
& \frac{1}{3} \sum_{j=1}^3 q_j = m_0
\label{Equation:q_Equality}
\end{align}
\end{subequations}
where 
\begin{align}
  m_0\equiv \sqrt{\frac{8}{27}} \,.
  \label{Equation:m0}
\end{align}

Consider the centered variables,
\begin{align}
    z_j=q_j-m_0 \,,
\label{Equation:z_j}
\end{align}
hence $q_j=z_j+m_0$.
The constraints can then be written as 
\begin{subequations}
\begin{align}
& 
z_j\geq -m_0 \,,\;
\mu\leq \frac{1}{3}      
\\
& \frac{1}{3} \sum_{j=1}^3 z_j = 0 \,.
\label{Equation:y_sqrt_reduced}
\end{align}
\end{subequations}

Let
\begin{align}
    \lambda&= \frac{1}{3}\sum_{j=1}^3 z_j^2 =\frac{\norm{\bm{z}}_2^2}{3} \,.
    \label{Equation:lambda}
\end{align}
Next, we would like to estimate $\mu$ and $\sigma^2$ in terms of $\lambda$.
\subsubsection{First-moment analysis}

Since $t_j=q_j^2=(z_j+m_0)^2$,
\begin{align}
    \mu% 
    &= \frac{1}{3}\left(\sum_{j=1}^3 z_j^2+2m_0\sum_{j=1}^3 z_j+3m_0^2 \right)
    \nonumber\\
    &= \lambda+m_0^2 
    \nonumber\\
    &= \lambda+\frac{8}{27} 
    \,.
    \label{Equation:mu_lambda}
\end{align}
Hence, 
the constraint in \eqref{Equation:y_mu_Constraint} implies
\begin{align}
    0\leq \lambda\leq \frac{1}{27} \,.
    \label{Equation:lambda_bound}
\end{align}

Furthermore,
the objective function is bounded by
\begin{align}
    g(\bm{t})&\leq
   \frac{8}{9} \log 3
   -3\left[2\sigma^2-
    \lambda \log 3\right]
    \label{Equation:Objective_Bound_sigma_1}
\end{align}
by \eqref{Equation:Objective_Bound_mu_sigma} and \eqref{Equation:mu_lambda}.
We would like to show that the expression in the square brackets is nonnegative. 

\subsubsection{Variance bound}
First observe that 
\begin{align}
 q_j^2-\mu
    &= (z_j+m_0)^2-\mu &&\text{by \eqref{Equation:z_j}}
    \nonumber\\
    &= (z_j+m_0)^2-\lambda-\frac{8}{27}  &&\text{by \eqref{Equation:mu_lambda}}
            \nonumber\\
    &= z_j^2  +2m_0 z_j -\lambda  &&\text{by \eqref{Equation:m0}.} \nonumber
\end{align}
Plugging this into \eqref{Equation:sigma2_y}, we obtain
\begin{align}
    &\sigma^2 = \frac{1}{3}\bigg[\sum_{j=1}^3  \left(  z_j^2  +2m_0 z_j -\lambda+\frac{1}{3}  \right)^2-\frac{1}{3}
    \bigg] 
    \nonumber\\
    &= \frac{1}{3}\bigg[\sum_{j=1}^3  \left(  z_j^2  +2m_0 z_j -\lambda  \right)^2
    \nonumber\\&\phantom{=}
    +2\cdot \frac{1}{3}\sum_{j=1}^3  \left(  z_j^2  +2m_0 z_j -\lambda  \right)
    +3\cdot\frac{1}{9}-\frac{1}{3}
    \bigg] 
        \nonumber\\
    &= \frac{1}{3}\bigg[ 
    \norm{ \bm{z}^2  +2m_0 \bm{z} -\lambda\bm{1}  }_2^2
    +\frac{2}{3}\left(3\lambda+2m_0\cdot 0-3\lambda \right)
    \bigg] \,,
\end{align}
where we have defined $\bm{z}^2\equiv (z_j^2)$.
Hence, 
\begin{align}
    \sigma 
    &= \frac{1}{\sqrt{3}}
    \norm{ 2m_0 \bm{z}+ \left( \bm{z}^2   -\lambda\bm{1} \right)  }_2
     \,.
    \label{Equation:sigma2_z}
\end{align}
Then, by the reverse-triangle inequality,
\begin{align}
   \sqrt{3} \sigma 
    &\geq % 
    2m_0\norm{  \bm{z}  }_2
    -
    \norm{  \bm{z}^2   -\lambda\bm{1}   }_2
    \nonumber\\
     &= 
    2m_0\sqrt{3\lambda}
    -
    \norm{  \bm{z}^2   -\lambda\bm{1}   }_2
     \label{Equation:sigma_z_bound_0}
\end{align}
(see \eqref{Equation:lambda}).

\subsubsection{Fourth-moment identity}
We use the following  identity:
    \begin{align}
        \sum_{j=1}^3 z_j^4 &=
        \frac{1}{2}\left( \sum_{j=1}^3 z_j^2 \right)^2
    \end{align}
We show in Appendix~\ref{Appendix:Fourth_Moment} that this identity holds for every zero-sum vector $\bm{z}\in\mathbb{R}^3$ (see Lemma~\ref{Lemma:Fourth_Moment}).
Equivalently,
\begin{align}
        \sum_{j=1}^3 z_j^4 &=
        \frac{1}{2}(3\lambda)^2 \,.
    \end{align}
Then,
\begin{align}
     \norm{  \bm{z}^2   -\lambda\bm{1}   }_2^2
     &= \sum_{j=1}^3 (z_j^2-\lambda)^2
     \nonumber\\
     &= \sum_{j=1}^3 z_j^4 -2\lambda \sum_{j=1}^3 z_j^2+3\lambda^2
     \nonumber\\
     &=9\left(\frac{1}{2}\lambda^2-\frac{2}{3}\lambda^2+\frac{1}{3}\lambda^2\right)
     \nonumber\\
     &=\frac{(3\lambda)^2}{6} \,.
\end{align}
Hence, 
\begin{align}
     \norm{  \bm{z}^2   -\lambda\bm{1}   }_2
     &=\frac{3\lambda}{\sqrt{6}} \,.
\end{align}
Substituting this into our bound on $\sqrt{3}\sigma$ in \eqref{Equation:sigma_z_bound_0}, we have
\begin{align}
    \sqrt{3}\sigma 
    &\geq        
    \sqrt{3\lambda}\cdot \left(
    2 m_0
    -
    \frac{\sqrt{3\lambda}}{\sqrt{6}}
    \right)
     \,.
     \label{Equation:sigma_z_bound}
\end{align}
We notice that
\begin{align}
    2 m_0=2\cdot \sqrt{\frac{8}{27}}% 
    = \frac{4\sqrt{6}}{9} \,.
\end{align}
Since 
$\sqrt{3\lambda}\leq \frac{1}{3}$ by
\eqref{Equation:lambda_bound},
\begin{align}
    \frac{\sqrt{3\lambda}}{\sqrt{6}}\leq \frac{1}{3\sqrt{6}}
    =  \frac{\sqrt{6}}{18}
\end{align}
thus,
\begin{align}
    \sigma 
    &\geq           
     \sqrt{\lambda}\cdot 
    \frac{7\sqrt{6}}{18}
     \,.
     \label{Equation:sigma_lambda_bound_1}
\end{align}
Hence,
$% 
      2\sigma^2 
    \geq                     \lambda\cdot 
    \frac{49}{27}
$. % 
At the same time, since 
$3^3<2^5$, we have 
$\log 3<\frac{5}{3}% 
<\frac{49}{27}$. Hence, 
\begin{align}
      2\sigma^2 
    &\geq                     \lambda\cdot 
    \log 3
    \,.
     \label{Equation:sigma_lambda_bound_4}
\end{align}
Equivalently,
$% 
      2\sigma^2 
    -  3\lambda\log 3 \geq 0
$, from which we can infer the desired upper bound: % 
 \begin{align}
    g(\bm{t})&\leq \frac{8}{9}\log 3
    \label{Equation:Objective_Bound_mu_sigma_1}
\end{align}
by \eqref{Equation:Objective_Bound_sigma_1}.
It follows that 
the capacity of the zero-delay relay channel without assistance satisfies
\begin{align}
    C_{\text{unassisted}}^{(0)}\leq \frac{8}{9}\log 3
    \,.
\end{align}

\subsection{Given Entanglement}
\label{Subsection:ET_Capacity}
The capacity of the zero-delay relay channel with entangled transmitters is given by 
\begin{align}
    C_{\text{ET}}^{(0)}&=\log 3
    \nonumber\\
    &\simeq 1.5849
\end{align}
Achievability is straightforward.
At every time instance, Alice can send a message 
$m_i\in\{1,2,3\}$ reliably. Alice and the relay simply play the magic-square game with respect to the inputs 
$q_0=m_i$ and $q_1=y_{r,i}$,
they measure $a_0$ and $a_1$, and then send
$x_i=(q_0,a_0)$ and $x_{r,i}=a_1$, respectively. 
Since the quantum strategy wins the game with certainty, Bob receives $y_i=q_0=m_i$ perfectly. 
The converse part is immediate as well, as we  achieve the full-cooperation rate in this case. 
This shows that entanglement shared between the sender and relay's transmitters provides a rate increase of at least $12.5\%$.
\qed

\section{Summary and Discussion}
\label{Section:Summary}

\subsection{Summary}
We studied classical relay channels in which the sender and relay share entanglement. The effect of this resource depends on the relay’s encoding delay. For a unit-delay relay, the standard cutset upper bound remains valid with entanglement assistance. This rules out a capacity gain for classes of relay channels in which the bound is tight, such as degraded and semi-deterministic relay channels. For a zero-delay relay, we constructed a channel based on the magic-square game.
Unlike the sum-rate separation established in previous work \cite{LeditzkyAlhejjiLevinSmith:20p,SeshadriLeditzkySiddhuSmith}, our example shows an entanglement-assisted gain in the communication rate of a single sender through a relay.
On this channel, entanglement violates the standard cutset bound and enables a rate increase of at least $12.5\%$.

The distinction arises from when the relay can use its observation. Under strictly causal encoding, the relay’s transmission at time $i$ cannot depend on $Y_{r,i}$. Under causal encoding, it can. Our example shows that this difference matters when the sender and relay share entanglement: the standard cutset bound for the unit-delay model need not bound the rates achievable in the zero-delay model.

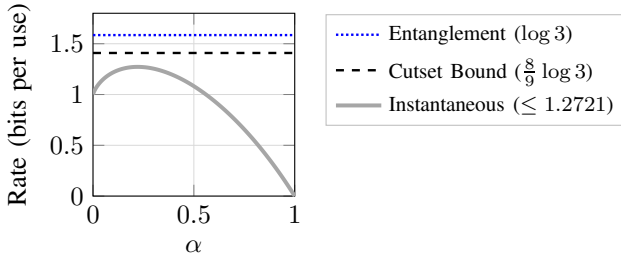
\begin{figure}
    \centering

\begin{tikzpicture}
  \begin{axis}[
   width=4.25cm,       % 
   height=4cm,      % 
    xlabel={$\alpha$},
    ylabel={Rate (bits per use)},
    xmin=0, xmax=1.0,
    ymin=0, ymax=1.8,
    xtick={0,  0.5,  1.0},
    ytick={0, 0.5, 1.0, 1.5},
    grid=both,
    grid style={line width=0.2pt, gray!30},
    axis lines=box,
    legend style={
      at={(2.65, 0.995)},
      anchor=north east,
      font=\footnotesize,
      fill=white,
      fill opacity=0.92,
      draw=gray!50,
      legend cell align=left,
      row sep=1.5pt
    },
    declare function={
      h2(\x) = (\x <= 0 ? 0 : -\x*(ln(\x)/ln(2)) - (1-\x)*(ln(1-\x)/ln(2)));
      d0 = h2(2/9) + (2/9); 
      f(\x) = h2(1-(7*\x)/9) + 1 - (7*\x)/9 - d0*\x;
      c(\x) = (8/9)*(ln(3)/ln(2));
      entangled(\x) = (ln(3)/ln(2));
    }
  ]

    \addplot[
      domain=0:1,
      samples=2,
      densely dotted,
      line width=0.9pt, % 
      blue
    ] {entangled(x)};
    \addlegendentry{Entanglement ($\log 3$)}% 

    \addplot[
      domain=0:1,
      samples=2,
      dashed,
      line width=0.9pt,
      black
    ] {c(x)};
    \addlegendentry{Cutset Bound ($\frac89\log 3$)}% 

    \addplot[
      domain=0.00001:1,
      samples=300,
      line width=1.5pt,
      gray!70!
    ] {f(x)};
    \addlegendentry{Instantaneous  ($\leq 1.2721$)}

  \end{axis}
\end{tikzpicture}

    \caption{Achievable rate for the row distribution $Q_1\sim \left(\frac{1-\alpha}{2},\frac{1-\alpha}{2},\alpha\right)$. }
    \label{Figure:Achievable_rate_alpha}
\end{figure}
\subsection{Classical Achievable Rate}

We may also derive achievable rates for the classical setting, without assistance.
Based on the instantenous-relaying lower bound, the rate 
$R=I(X;Y)$ is achievable for every input distribution
\cite{ElGamalHassanpourMammen:07p}.
Consider the magic-square game in the classical setting. 
If the questions are uniform, then the following deterministic strategy is optimal:
\begin{subequations}
    \label{Equation:Achievability_Classical_Strategy}
\begin{align}
    a_0(1)&=\begin{pmatrix}
        0 & 0& 0
    \end{pmatrix}
    \\
    a_0(2)&=\begin{pmatrix}
        0 & 0& 0
    \end{pmatrix}
    \\
    a_0(3)&=\begin{pmatrix}
        1 & 1& 0
    \end{pmatrix}
\intertext{and}
    a_1(q_1)&=\begin{pmatrix}
        0 \\ 0\\ 1
    \end{pmatrix} \;\forall q_1 
\end{align}
\end{subequations}
(see Table~\ref{tab:classical-magic-square}).
In this case, if $Q_0\in \{1,2\}$, the players win the game. 
Otherwise, if $Q_0=3$, they win in two out of three cases. 

However, we can % 
increase the winning probability if we set $Q_0$ to be non-uniform, with $p_{Q_0}(1)=p_{Q_0}(2)=\frac{1-\alpha}{2}$ and $p_{Q_0}(3)=\alpha$.
This yields the achievable rate in  Figure~\ref{Figure:Achievable_rate_alpha}, where the instantaneous-relaying rate is indicated by the gray solid  line.
The derivation of the lower bound is provided in Appendix~\ref{Appendix:Instanteneous_Relaying_Bound}.
Clearly, there is a substantial gap between this lower bound and our cutset upper bound, 
$\frac89 \log 3\simeq 1.409$, as  indicated by the dashed line.
The capacity without entanglement satisfies
\begin{align}
    1.2721\leq  C_{\text{unassisted}}^{(0)}\leq 1.4089 \,.
\end{align}

The capacity with entangled transmitters 
$C_{\text{unassisted}}^{(0)}=\log 3\simeq 1.5849$
is shown as the blue dotted line in Figure~\ref{Figure:Achievable_rate_alpha}.
Based on these values, the rate increase lies between $12.5\%$ and $24.6\%$.

\subsection{Converse Methods}
Our converse differs from the analyses developed for game-based MACs. For the magic-square MAC, entanglement between the senders enables a sum rate of $2\log 3$. The unassisted sum capacity remains unknown. Leditzky et al.~\cite{LeditzkyAlhejjiLevinSmith:20p} obtained an upper bound by relating the output entropy to the losing probability. This establishes a sum-rate increase of at least $1.05\%$. Seshadri et al.~\cite{SeshadriLeditzkySiddhuSmith} formulated a constrained optimization problem and introduced a relaxation. Their analytical bound establishes an increase of at least $4.7\%$.

For our zero-delay relay channel, the capacity with entanglement between the sender and relay is $\log 3$. We initially explored an adaptation of the approach of Leditzky et al.~\cite{LeditzkyAlhejjiLevinSmith:20p}. This calculation, not included here, yielded the upper bound $C^{(0)}_{\mathrm{unassisted}}\leq 1.5622$, establishing a capacity increase of at least $1.45\%$. 

The converse presented here takes a different route. The optimization perspective of Seshadri et al.~\cite{SeshadriLeditzkySiddhuSmith} motivated us to consider a relaxation. Our relaxation retains the parity relations among classical responses to different magic-square questions. An entropy inequality then bounds the resulting objective analytically. This yields the explicit upper bound $\frac{8}{9}\log 3$ and establishes a capacity increase of at least $12.5\%$. Thus, the connection to Seshadri et al.~\cite{SeshadriLeditzkySiddhuSmith} lies in the optimization perspective, while our bound relies on the specific structure of classical magic-square strategies.

\subsection{Future Directions}

These results invite further study of how network structure affects the value of shared entanglement.
Several questions remain open. For the magic-square relay channel, determining the exact unassisted capacity would establish the full extent of the entanglement advantage. More broadly, a systematic treatment of game-based relay channels could clarify how properties of the underlying game affect communication rates. Extending this framework to other network settings could reveal further roles for shared entanglement. Another direction is to explore applications of these channels in cryptographic protocols based on nonlocal games.

\section*{Acknowledgments}
The authors % 
thank Gerhard Kramer (TUM) for useful discussions.
This work was supported by  
ISF, n. 939/23, 2691/23, 
OMC, n. 86160946, DIP, n.  2032991,  
Chaya % 
Chair, n. 8776026, 
 Helen Diller Quantum Center, n. 2033613, and
 VATAT % 
 Program for Quantum Science and Technology, n. 86636903.
The authors acknowledge the assistance of ChatGPT 5.6 in this work. The authors initiated a series of discussions on entanglement-assisted relay settings and the magic-square example. After the authors formulated the optimization problem arising from the cutset bound, ChatGPT 5.6 suggested the use of a quadratic entropy bound to sharpen the classical analysis. ChatGPT 5.6 was also used for proofreading. The authors developed the proofs and accept full responsibility for the accuracy and originality of the results.

\begin{appendices}
\numberwithin{theorem}{section}

\setcounter{theorem}{0}
    \section{Proof of Lemma~\ref{Lemma:Distinct_Coordinates}}
    \label{Appendix:Distinct_Coordinates}

\subsection{Boundary Exclusion}
We first claim that a minimizer cannot lie on the simplex boundary.
\begin{lemma}
\label{Lemma:Boundary_exclusion}
    If $\bm{p}$ minimizes $\Delta$, then $p_j>0$ for $j\in\{1,2,3\}$.
\end{lemma}

\begin{proof}[Proof of Lemma~\ref{Lemma:Boundary_exclusion}]    
Assume to the contrary that $\bm{p}$ is a minimizer on the boundary as e.g., $p_3=0$.  
Then, at least one of the other coordinates, $p_1$ and $p_2$, is positive. 
Suppose $p_1>0$. Now, consider a new probability vector,
\begin{align}
    \bm{p}_\varepsilon=(p_1-\varepsilon,p_2,\varepsilon)
\end{align}
for $\varepsilon\in [0,p_1)$. 

Now, we estimate the derivative of $F(\varepsilon)=\Delta(\bm{p}_\varepsilon)$ at $\varepsilon\to 0^+$. Observe that 
\begin{align}
    F(\varepsilon)&=
    (p_1-\varepsilon) \log(3(p_1-\varepsilon))+p_2\log(3p_2)
    +\varepsilon\log(3\varepsilon)
    \nonumber\\&\phantom{=}
    -2\left( (p_1-\varepsilon)^2+p_2^2+\varepsilon^2 -\frac{1}{3}
    \right)
    \nonumber\\
    &=
    \phi(\varepsilon)+p_2\log(3p_2)
    +\varepsilon\log(3\varepsilon)
    \nonumber\\&\phantom{=}
    -2\left( (p_1-\varepsilon)^2+p_2^2+\varepsilon^2 -\frac{1}{3}
    \right)
    \intertext{where}
    \phi(\varepsilon)&= (p_1-\varepsilon) \log(3(p_1-\varepsilon))
    \,.
\end{align}
Hence,
\begin{align}
    &F(\varepsilon)-F(0)
    \nonumber\\
    &=
        \phi(\varepsilon)+p_2\log(3p_2)
    +\varepsilon\log(3\varepsilon)
    \nonumber\\&\phantom{=}
    -2\left( (p_1-\varepsilon)^2+p_2^2+\varepsilon^2 -\frac{1}{3}
    \right)
    \nonumber\\&\phantom{=}
    -\left[ \phi(0)+p_2\log(3p_2)
     \right]
     \nonumber\\&\phantom{=}
    +2\left( p_1^2+p_2^2 -\frac{1}{3}
    \right)    
    \nonumber\\
    &=
        \phi(\varepsilon)+p_2\log(3p_2)
    +\varepsilon\log(3\varepsilon)
    \nonumber\\&\phantom{=}
    -2\left( p_1^2-2p_1\varepsilon+\varepsilon^2 +p_2^2+\varepsilon^2 
    \right)
    \nonumber\\&\phantom{=}
    -\left[ \phi(0)+p_2\log(3p_2)
     \right]
     \nonumber\\&\phantom{=}
    +2\left( p_1^2+p_2^2 
    \right) 
        \nonumber\\
    &=
        \phi(\varepsilon)
    +\varepsilon\log(3\varepsilon)
    -2\left( -2p_1\varepsilon+2\varepsilon^2  
    \right)
    - \phi(0) \,.
    \end{align}
Rearranging the terms and dividing by $\varepsilon$, we have 
\begin{align}
    \frac{F(\varepsilon)-F(0)}{\varepsilon}&=
    \frac{\phi(\varepsilon)- \phi(0)}{\varepsilon}
    +4 \left( p_1-\varepsilon
    \right)
    +\log(3\varepsilon) \,.
\end{align}
Taking the limit of $\varepsilon\to 0^+$,
the first term becomes the right derivative, $\phi'(0^+)$.
The derivative is given by
\begin{align}
    \phi'(\varepsilon)
    &=-\log(3(p_1-\varepsilon))-
    \frac{1}{\ln 2}
\end{align}
for $\varepsilon\in (0,p_1)$. 
Then, we obtain
\begin{align}
    F'(0^+)&=\lim_{\varepsilon\to 0^+} \frac{F(\varepsilon)-F(0)}{\varepsilon}
    \nonumber\\
    &=-\log(3p_1)-
    \frac{1}{\ln 2}
    +4 p_1
    +\lim_{\varepsilon\to 0^+}\log(3\varepsilon)
    \nonumber\\
    &=-\infty \,.
\end{align}

It follows that for sufficiently small $\varepsilon>0$, 
\begin{align}
    F(\varepsilon)<F(0) \,.
\end{align}
Hence, the boundary point $\bm{p}_{\varepsilon=0}=\bm{p}$ cannot be a minimum. % 
We deduce that our initial assumption is false, and thus, 
every $\bm{p}$ that minimizes $\Delta(\bm{p})$ has full support. 
This completes the proof of Lemma~\ref{Lemma:Boundary_exclusion}.
\end{proof}

\subsection{First-Order Necessary Conditions}

Next,
we analyze first-order necessary conditions for optimality \cite[Sec. 12.3]{NocedalWrite:06b}.
Consider the Lagrangian,
\begin{align}
    \mathcal{L}(\bm{p},\lambda,\mu)&=
    \Delta(\bm{p})+\mu\left( \sum_{j=1}^3 p_j-1 \right)-\sum_{j=1}^3 \lambda_j p_j 
    \,,
\end{align}
with $\mu$, $\lambda_j$, $j\in \{1,2,3\}$, as Lagrange multipliers. 
The KKT conditions are
\begin{subequations}
   \begin{align}
   & p_j\geq 0 \,,\; \sum_{j'=1}^3 p_{j'}=1 \\
   & \lambda_j\geq 0 \,, \label{Equation:Positivity_Constraint}\\
   & \lambda_j p_j=0 \,, \label{Equation:Complementary_Slackness}
    \intertext{and}
   & \frac{\partial \mathcal{L}}{\partial p_j}=0
   \label{Equation:Derivative_Constraint}
\end{align} 
for $j\in\{1,2,3\}$.
\end{subequations}
We also need to verify that the assumptions of the first-order necessary coniditions \cite[Th. 12.1]{NocedalWrite:06b} are met.
Indeed, the objective function and constraints are all smooth.
Furthermore, by Lemma~\ref{Lemma:Boundary_exclusion}, 
our optimization has a single constraint with active equality. 
Hence, linear independence of the active equality constraints trivially holds. This precisely means that the property of linear independence constraint qualification (LICQ) holds. 
Thus, every minimizer must satisfy the KKT conditions above  \cite[Th. 12.1]{NocedalWrite:06b}.

Based on Lemma~\ref{Lemma:Boundary_exclusion},
the complementary-slackness constraint, \eqref{Equation:Complementary_Slackness},
implies
\begin{align}
    \lambda_1=\lambda_2=\lambda_3=0 \,.
\end{align}
Then,
\begin{align}
    \frac{\partial \mathcal{L}}{\partial p_j}&=
    \frac{\partial \Delta}{\partial p_j}+\mu
    \nonumber\\
    &=\log(3p_j)+\frac{1}{\ln 2}-4p_j+\mu
    \nonumber\\
    &=\log3+\log p_j+\log e-4p_j+\mu
    \,.
\end{align}
Setting this to zero gives
\begin{align}
\log(p_j)-4p_j=-\mu-\log( 3e)
\label{Equation:KKT_Consequence}
\end{align}
for $j\in \{1,2,3\}$.
This proves part~1 of Lemma~\ref{Lemma:Distinct_Coordinates}.

\subsection{Distinct Coordinates}
We now prove part~2 of Lemma~\ref{Lemma:Distinct_Coordinates}.
Based on the derivation above, % 
if $\bm{p}$ is a minimizer, then 
\begin{align}
\log(x)-4x=\text{Const.}
\label{Equation:KKT_Consequence_alpha}
\end{align}
must hold for all coordinates, i.e., $x=p_1$, $x=p_2$, and $x=p_3$.

    Consider the function
    \begin{align}
        \theta(x)=\log x-4x
    \end{align}
    on $(0,1]$. Then, by direct computation, the first and second derivatives are:
    \begin{align}
        \theta'(x)&=\frac{1}{x\ln2}-4 \,,\\
        \theta''(x)&=-\frac{1}{x^2\ln2} <0 \,.
    \end{align}
    Thus, $\theta$ is strictly concave. 
    Specifically, $\theta(x)$ increases for $x<\frac{1}{4\ln 2}$, and then decreases for $x>\frac{1}{4\ln 2}$. It follows that  a horizontal line can intersect its graph at no more than two points. Hence, there can be at most two distinct solutions to Equation~\eqref{Equation:KKT_Consequence_alpha}.
    This concludes the proof of Lemma~\ref{Lemma:Distinct_Coordinates}.
    \qed

 \setcounter{theorem}{0}
\section{Fourth-Moment Identity}
\label{Appendix:Fourth_Moment}
We show the following identity. 
\begin{lemma}
\label{Lemma:Fourth_Moment}
    For every zero-sum vector $\bm{z}\in\mathbb{R}^3$,
    \begin{align}
        \sum_{j=1}^3 z_j^4 &=
        \frac{1}{2}\left( \sum_{j=1}^3 z_j^2 \right)^2
    \end{align}
\end{lemma}
\begin{proof}
First, we note that
    \begin{align}
        \left( \sum_{j=1}^3 z_j^2 \right)^2
        =\sum_{j=1}^3 z_j^4 +2\sum_{j<\ell} z_j^2 z_\ell^2
        \,.
        \label{Equation:Square_sum}
    \end{align}
    Since the vector has a zero sum,
    \begin{align}
        0&= (z_1+z_2+z_3)^2
        \nonumber\\
        &= \sum_{j=1}^3 z_j^2+ 2(z_1 z_2+z_1 z_3+z_2 z_3)
    \end{align}
    hence, 
    \begin{align}
        z_1 z_2+z_1 z_3+z_2 z_3
        &= -\frac{1}{2} \sum_{j=1}^3 z_j^2 \,.
    \end{align}
    Squaring this identity yields
    \begin{align}
      \sum_{j<\ell} (z_j z_\ell)^2         +2z_1 z_2 z_3(z_1 +  z_2 +  z_3)
         =\frac{1}{4} \left(\sum_{j=1}^3 z_j^2 \right)^2 
    \end{align}
    or, equivalently,
    \begin{align}
        \sum_{j<\ell} z_j^2 z_\ell^2
        = \frac{1}{4} \left(\sum_{j=1}^3 z_j^2 \right)^2 \,,
    \end{align}
    as $z_1+z_2+z_3=0$ by assumption.
    Plugging this into \eqref{Equation:Square_sum}, we have
    \begin{align}
        \left( \sum_{j=1}^3 z_j^2 \right)^2
        =\sum_{j=1}^3 z_j^4 +2\cdot \frac{1}{4} \left(\sum_{j=1}^3 z_j^2 \right)^2
    \end{align}
    hence,
    \begin{align}
        \frac{1}{2}\left( \sum_{j=1}^3 z_j^2 \right)^2
        =\sum_{j=1}^3 z_j^4
    \end{align}
    as we were set to prove.
\end{proof}

\section{Instantaneous-Relaying Rate}
\label{Appendix:Instanteneous_Relaying_Bound}
We derive an achievable rate for the classical zero-delay relay channel, without assistance. 
Based on the instantenous-relaying lower bound \cite{ElGamalHassanpourMammen:07p} \cite [Sec. 16.9.2]{ElGamalKim:11b},
\begin{align}
    C_{\text{unassisted}}^{(0)}\geq \max_{p_{X}p_{X_r|Y_r}} I(X;Y)
    \,,
\end{align}
for
$(X,Y_r,X_r,Y)\sim
p_{X}(x) P_{Y_r|X}(y_r|x)\cdot
p_{X_r|Y_r}(x_r|y_r) P_{Y|XX_rY_r}(y|x,x_r,y_r)$.

Consider the magic-square game in the classical setting. 
If the questions are uniform, then the deterministic strategy in \eqref{Equation:Achievability_Classical_Strategy} is optimal.
In this case, if $Q_0\in \{1,2\}$, the players win the game. 
Otherwise, if $Q_0=3$, they win in two out of three cases. 
Nevertheless, we can % 
increase the winning probability if we set $Q_0$ to be non-uniform, with $p_{Q_0}(1)=p_{Q_0}(2)=\frac{1-\alpha}{2}$ and $p_{Q_0}(3)=\alpha$.
This yields % 
\begin{align}
    H(\widehat{Q}_0)=h_2\left( 1-\frac{7}{9}\alpha \right)+1-\frac{7}{9}\alpha \,.
\end{align}

We move to the conditional output entropy $H(\widehat{Q}_0|Q_0 A_0)=H(\widehat{Q}_0|Q_0)$.
Given $Q_0\in \{1,2\}$, the entropy is zero. % 
Whereas, given $Q_0=3$, the conditional entropy is 
\begin{align}
    b_0\equiv % 
    h_2\left( \frac{2}{9} \right)+\frac{2}{9} \,.
\end{align}
Averaging over $Q_0\in \{1,2,3\}$, we obtain
\begin{align}
    H(\widehat{Q}_0|Q_0)% 
    &= b_0 % 
    \alpha \,.
\end{align}

We deduce that  % 
\begin{align}
      I(Q_0 A_0;\widehat{Q}_0)
      &=h_2\left( 1-\frac{7}{9}\alpha \right)+1-\frac{7}{9}\alpha
     -b_0 \alpha 
\end{align}
is an achievable rate, for every $\alpha\in [0,1]$. % 
See Figure~\ref{Figure:Achievable_rate_alpha}, where the instantaneous-relaying rate is indicated by the solid line.
Through straightforward arithmetic, we find that the maximal value is attained for
$\alpha^*=\frac{9}{7}\left({1+2^{1+\frac{9}{7}b_0}}\right)^{-1}
$. This yields 
\begin{align}
     C_{\text{unassisted}}^{(0)} &\geq I(Q_0 A_0;\widehat{Q}_0)
                   \nonumber\\
     &=  \log\left( {2+2^{-\frac{9}{7}b_0}}  \right)
     \nonumber\\
     &\simeq 1.2721 % 
    \,.
\end{align}
\end{appendices}

{

\balance

\bibliography{references}

}

\end{document}